\pdfoutput=1

\documentclass[11pt,twoside,a4paper,cmspaper,final,collab]{cms-tdr}
\def\svnVersion{5257781-D}\def\svnDate{2020/09/19}\def\cmsCernNoTag{CERN-EP-2020-057}\def\cmsCernDate{\today}\def\cmsMessage{"Published in Physical Review D as \href{http://dx.doi.org/10.1103/PhysRevD.102.072001}{\doi{10.1103/PhysRevD.102.072001}.}"}

\begin{document}\cmsNoteHeader{HIG-18-021}

\hyphenation{had-ron-i-za-tion}
\hyphenation{cal-or-i-me-ter}
\hyphenation{de-vices}
\RCS$HeadURL: svn+ssh://svn.cern.ch/reps/tdr2/papers/HIG-18-021/trunk/HIG-18-021.tex $
\RCS$Id: HIG-18-021.tex 494080 2019-04-24 06:20:00Z rverma $

\newlength\cmsFigWidth
\ifthenelse{\boolean{cms@external}}{\setlength\cmsFigWidth{0.49\textwidth}}{\setlength\cmsFigWidth{0.65\textwidth}} 
\ifthenelse{\boolean{cms@external}}{\providecommand{\cmsLeft}{upper\xspace}}{\providecommand{\cmsLeft}{left\xspace}}
\ifthenelse{\boolean{cms@external}}{\providecommand{\cmsRight}{lower\xspace}}{\providecommand{\cmsRight}{right\xspace}}

\newcommand{\mjj}{\ensuremath{m_\text{jj}}\xspace}
\newcommand{\mt}{\ensuremath{m_{\PQt}}\xspace}
\newcommand{\mhp}{\ensuremath{m_{\PSHp}}\xspace}
\newcommand{\Bthb}{\ensuremath{\mathcal{B}(\PQt \to \PSHp\PQb)}\xspace}
\newcommand{\Bthbbar}{\ensuremath{\mathcal{B}(\PAQt \to \PSHm\PAQb)}\xspace}
\newcommand{\BHcs}{\ensuremath{\mathcal{B}(\PSHp \to \PQc\PAQs)}\xspace}
\newcommand{\BHtau}{\ensuremath{\mathcal{B}(\PSHp \to \PGtp\PGnGt)}\xspace}
\newcommand{\ejets}{$\Pe$\,+\,jets}
\newcommand{\mujets}{$\PGm$\,+\,jets}
\ifthenelse{\boolean{cms@external}}{\providecommand{\cmsTable}[1]{#1}}{\providecommand{\cmsTable}[1]{\resizebox{\textwidth}{!}{#1}}}
\newlength\cmsTabSkip\setlength{\cmsTabSkip}{1ex}

\cmsNoteHeader{HIG-18-021}

\title{Search for a light charged Higgs boson in the \texorpdfstring{$\PHpm \to \PQc\PQs$}{H(+/-) to cs}
    channel in proton-proton collisions at \texorpdfstring{$\sqrt{s}=13\TeV$}{sqrt(s)= 13 TeV}}

\date{\today}

\abstract{
A search is conducted for a low-mass charged Higgs boson produced
in a top quark decay and subsequently decaying into a charm and
a strange quark. The data sample was recorded in proton-proton
collisions at $\sqrt{s}=13\TeV$ by the CMS experiment at the LHC
and corresponds to an integrated luminosity of 35.9\fbinv. The search
is performed in the process of top quark pair production,
where one top quark decays to a bottom quark and a charged Higgs
boson, and the other to a bottom quark and a \PW boson. With
the \PW boson decaying to a charged lepton (electron or muon) and a
neutrino, the final state comprises an isolated lepton, missing
transverse momentum, and at least four jets, of which two are tagged
as \PQb jets. To enhance the search sensitivity, one of the jets
originating from the charged Higgs boson is required to satisfy a
charm tagging selection. No significant excess beyond standard model
predictions is found in the dijet invariant mass distribution. An
upper limit in the range 1.68--0.25\% is set on the branching fraction
of the top quark decay to the charged Higgs boson and bottom quark for
a charged Higgs boson mass between 80 and 160\GeV.
}
\hypersetup{%
pdfauthor={CMS Collaboration},%
pdftitle={Search for a light charged Higgs boson in the H(+/-) to cs channel in proton-proton collisions at sqrt(s) = 13 TeV},%
pdfsubject={CMS},%
pdfkeywords={CMS, physics, charged Higgs boson, MSSM}}
\maketitle

\section{Introduction}
\label{s:secIntro}
The discovery of the Higgs boson in 2012 by the ATLAS~\cite{Aad:2012tfa}
and CMS~\cite{Chatrchyan:2012ufa,Chatrchyan:2013lba} experiments at
the CERN LHC has given rise to a wide set of measurements to establish the 
nature of the discovered particle. The Higgs boson could be the first 
of many elementary scalars present in nature to be observed in the 
laboratory. Various extensions of the standard model (SM), such as the 
two Higgs doublet model (2HDM)~\cite{Branco:2011iw}, including
supersymmetry~\cite{Martin:1997ns,Golfand:1971iw,Wess:1974tw}, predict
multiple scalars as the remnants of an additional SU(2)$_L$ complex
doublet introduced to address some known limitations of the SM, such as
the origin of dark matter~\cite{Ade:2013zuv,Ade:2013sjv} and the
hierarchy problem~\cite{ArkaniHamed:1998rs}. After spontaneous symmetry
breaking, out of the eight degrees of freedom of the two Higgs
doublets, three are used to make the \PW and \PZ bosons massive,
leaving five physical scalar particles. Of these, two are
neutral Higgs bosons that are CP-even (scalar), one is neutral
and CP-odd (pseudoscalar), and the remaining two are charged Higgs
bosons (\PHpm).

The 2HDM can be classified into different categories depending on the
type of interaction of the two doublets with quarks and charged
leptons. For example, in the type II 2HDM, leptons and down-type 
quarks have Yukawa couplings to the first doublet, and up-type 
quarks couple to the second doublet. The nature of the Yukawa
coupling determines the branching fraction $\mathcal{B}$ of the charged Higgs boson
decays into different final states. We are interested in the search
for a low-mass ($\mhp < \mt$) charged Higgs boson in the decay
channel $\PSHp \to \PQc\PAQs$ (and its charge conjugate), whose
branching fraction can range up to 100\%, depending on the type of
Yukawa coupling. The latter is expressed in terms of the parameter
$\tan\beta=v_2/v_1$, where $v_1$ and $v_2$ are the vacuum expectation
values of the two Higgs doublets. In the minimal supersymmetric
standard model, this is the dominant decay channel for low
values of $\tan\beta$ for most of the mass range considered in this
analysis~\cite{Aoki:2009ha,Ma:1997up}. We assume that
$\BHcs = 100\%$.

\begin{figure*}[htp]
\centering
\includegraphics[width=0.8\textwidth]{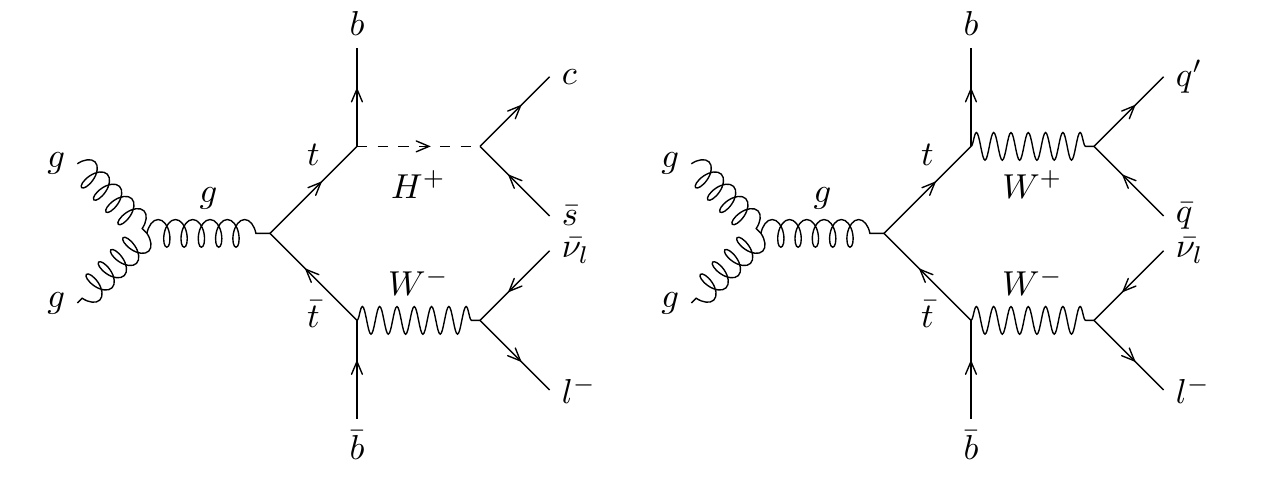}
\caption{Sample diagrams of \ttbar production via gluon-gluon
    scattering. The left plot shows the signal process in which the
    \ttbar pair decay products include a charged Higgs boson. The
    right plot shows the SM decay of a \ttbar pair in the
    semileptonic decay channel.}
\label{fig:feyn_diag_sig}
\end{figure*}

As illustrated in Fig.~\ref{fig:feyn_diag_sig}, in the signal process
for \PSHp production, one of the top quarks decays to $\PSHp\PQb$ and
the other to $\PWm\PAQb$, with \PSHm production proceeding by the
charge conjugate of this process. The principal SM background to this
search consists of \ttbar pair production where both top quarks decay
to a \PW boson and a \PQb quark. In this search, we consider the mode
where the $\PWp/\PSHp$ decays hadronically into a charm and strange antiquark, 
whereas the \PWm decays leptonically (in the \ttbar case, this is 
called the ``semileptonic'' decay channel); we define two channels 
depending on whether the lepton produced in the \PWm decay is a muon 
or an electron (events with tau leptons are not specifically considered,
but can be selected if the tau lepton decays into a muon or an electron).

There have been many earlier searches for charged Higgs bosons at LEP,
the Tevatron, and the LHC. At LEP, these were expected to be
dominantly produced by the process $\Pep\Pem \to \PSHp\PSHm$. Assuming
that \PSHp decays only to $\PQc\PAQs$ and $\PGtp\PGnGt$, \ie, the sum
of the branching factions $\BHtau +
\BHcs=1$, lower limits of 79.3 and
80.0\GeV were set on the charged Higgs boson mass at 95\% confidence
level (\CL) from individual
collaborations~\cite{Achard:2003gt, Heister:2002ev, Abdallah:2003wd}
and combined LEP data~\cite{Abbiendi:2013hk}, respectively.
Under a more general assumption $\BHtau +
\mathcal{B}(\PSHp \to \PQq\PAQq^\prime)=1$, a slightly less stringent
constraint of 76.3\GeV was obtained at 95\% \CL~\cite{Abbiendi:2008aa}.

Limits on charged Higgs boson production at hadron colliders were set
by the Tevatron and LHC experiments, assuming the production mode
$\PQt \to \PSHp\PQb$. The CDF
Collaboration~\cite{Aaltonen:2009ke} set a 95\% \CL upper limit on the
branching fraction \Bthb of 10--30\% for a
charged Higgs boson mass lying in the range 60--150\GeV, assuming that
\PSHp decays only to $\PQc\PAQs$. Similar limits were obtained
by the D0 Collaboration~\cite{Abazov:2009aa}. Using 8\TeV data, the
ATLAS~\cite{Aad:2014kga} and CMS~\cite{Khachatryan:2015qxa}
Collaborations set an upper limit at 95\% \CL on the product
$\Bthb \BHtau$ of 1.3--0.23\% and 1.2--0.13\%, respectively, for a charged Higgs boson
mass in the range 80--160\GeV. A search for a charged Higgs boson
decaying into $\PQc\PAQs$ was performed with 7 (8)\TeV data by the
ATLAS (CMS) Collaboration, which set an upper limit at 95\% \CL on
\Bthb in the range $<$5.1 (6.5--1.2)\%
for a charged Higgs boson mass between 90 and
160\GeV~\cite{Aad:2013hla,Khachatryan:2015uua}. The CMS Collaboration
also performed a search for a charged Higgs boson in the $\PSHp \to \PQc\PAQb$
channel and put the most stringent upper limit at 95\% \CL on
\Bthb in the range 0.8--0.5\% for a charged
Higgs boson mass in the range 90 to 150\GeV~\cite{Sirunyan:2018dvm}.

At 13\TeV, the ATLAS and CMS Collaborations have performed several
searches for charged Higgs bosons in different
search channels such as $\PSHp \to \PGtp\PGn$, $\PSHp \to \PQt\PAQb$,
$\PSHp \to \PWp\PZ$, and $\PSHp \to \PWp\PSA$~\cite{Aaboud:2016dig,
Aaboud:2018gjj,Aaboud:2018cwk,Sirunyan:2019zdq,Sirunyan:2017sbn,
Sirunyan:2019hkq}.
The most stringent upper limit on $\sigma (\Pp\Pp\to\PQt\PSHp +\text{X})
\mathcal{B}(\PSHp \to \PGtp\PGn)$ at 95\% \CL is
4.2--0.0025\unit{pb} for a charged Higgs boson mass in the range from
90 to 2000\GeV from ATLAS~\cite{Aaboud:2018gjj}. The ATLAS Collaboration 
has also set an upper limit at 95\% \CL on $\sigma (\Pp\Pp\to\PQt\PSHp +\text{X})
\mathcal{B}(\PSHp \to \PQt\PAQb)$ in the range 9.6--0.01\unit{pb} for
a charged Higgs boson mass in the range 200 to 3000\GeV~\cite{Aaboud:2018cwk}.
Low values of $\tan\beta < 1$ are excluded for a charged Higgs boson
mass up to 160\GeV by both ATLAS and CMS~\cite{Aaboud:2018gjj,Sirunyan:2019hkq}.

This paper is organized as follows. A brief introduction about the CMS
detector is given in Section~\ref{s:secCMS}, followed by the description
of collision data and simulated samples in Section~\ref{s:secDataMC}.
The reconstruction of various physics objects such as the primary 
vertex, muons, electrons, jets, and missing transverse momentum are 
described in Section~\ref{s:secReco}. The event selection and 
background estimation method are explained in Section~\ref{s:secEvtSel}. 
The kinematic fitting and categorization of events based on charm jet 
tagging is discussed in Section~\ref{s:secKF}. The systematic and 
statistical uncertainties are described in Section~\ref{s:secSys}. The 
results are presented in Section~\ref{s:secResults}, followed by the 
summary in Section~\ref{s:secConcl}.

\section{The CMS detector}
\label{s:secCMS}
The central feature of the CMS apparatus is a superconducting solenoid
of 6\unit{m} internal diameter, providing a magnetic field of
3.8\unit{T}. Within the solenoid volume are a silicon pixel and strip
tracker, a lead tungstate crystal electromagnetic calorimeter (ECAL),
and a brass and scintillator hadron calorimeter, each composed
of a barrel and two endcap sections. The silicon pixel and tracker
detectors identify the trajectory of charged particles and accurately
measure their transverse momentum \pt up to pseudorapidity $\abs{\eta} \leq 2.5$.  Forward
calorimeters extend the $\eta$ coverage provided by the barrel
and endcap detectors. Segmented calorimeters provide sampling of
electromagnetic and hadronic showers up to $\abs{\eta} \leq 5$. Muons are
detected in gas-ionization chambers embedded in the steel flux-return
yoke outside the solenoid, in the range of $\abs{\eta} \leq 2.4$. 

Events of interest are selected using a two-tiered trigger
system~\cite{Khachatryan:2016bia}. The first level (L1), composed of
custom hardware processors, uses information from the calorimeters and
muon detectors to select events at a rate of around 100\unit{kHz}
within a time interval of less than 4\mus. The second level, known as
the high-level trigger (HLT), consists of a farm of processors running
a version of the full event reconstruction software optimized for fast
processing, and reduces the event rate to around 1\unit{kHz} before
data storage. A more detailed description of the CMS detector,
together with a definition of the coordinate system used and the
relevant kinematic variables can be found in Ref.~\cite{Chatrchyan:2008zzk}.

\section{Data and simulation}
\label{s:secDataMC}
The data used for the analysis were collected with the CMS detector in
2016, in proton-proton ($\Pp\Pp$) collisions at $\sqrt{s} = 13\TeV$,
and correspond to an integrated luminosity of 35.9\fbinv.

As shown in Fig.~\ref{fig:feyn_diag_sig}, the charged Higgs boson is
assumed to decay into $\PQc\PAQs$ or $\PAQc\PQs$ only. As a result, in
the final state, there will be four jets (two \PQb jets, one \PQc jet,
one \PQs jet), one lepton (\Pgm or \Pe; \Pgt is not considered in this
analysis), and missing transverse momentum (\ptmiss), which is
attributed to the neutrino. The SM processes that give the same final
states (four jets\,+\,one lepton\,+\,missing transverse momentum) are
considered as background processes for this analysis. Signal and
background processes are modeled using simulated samples, generated
using the \MGvATNLO v2.3.3~\cite{Alwall:2014hca} and \POWHEG
v2.0~\cite{Frixione:2007vw, Nason:2004rx, Alioli:2010xd, Ball:2017nwa}
generators at parton level, with the NNPDF 3.0~\cite{Ball:2017nwa}
parton distribution functions (PDFs), with the order matching that in
the matrix element calculations. In all cases, these parton-level
events are hadronized using \PYTHIA 8.212~\cite{Sjostrand:2014zea}
with the CUETP8M1 underlying event tune~\cite{Khachatryan:2015pea} and
then passed to \GEANTfour~\cite{Agostinelli:2002hh} for simulation of
the CMS detector response. Finally, the events are reconstructed after
complete detector simulation using the same reconstruction process as
for data.

The SM \ttbar process is an irreducible background, and represents
the largest contribution, about 94\% of the total expected background
in the signal region. The parton-level SM inclusive \ttbar events, 
which have contributions from semileptonic, fully leptonic, and 
fully hadronic decay modes, are generated at next-to-leading order (NLO) 
using \POWHEG. The next-to-NLO cross section for \ttbar is calculated 
to be $\sigma_{\ttbar} = 832 \pm^{20}_{29}(\text{scale}) \pm 35\,
(\text{PDF} + \alpS)\unit{pb}$~\cite{Beneke:2011mq}. The top quark
mass in the simulated samples is taken to be 172.5\GeV.

The charged Higgs boson signal samples are generated using \MGvATNLO
at leading order (LO). Only \PSHp samples are generated, and \PSHm
production is assumed to be the same. The signal sample is generated
for several mass points in the range of 80 to 160\GeV (80, 90, 100,
120, 140, 150, 155, and 160\GeV). The generated cross section for the
signal is taken to be $0.21 \sigma_{\ttbar}$, where the factor of
0.21 is the branching fraction of $\PWm \to \ell^- \PAGn_\ell$ (where
$\ell = \Pgm$ or \Pe, neglecting the small contribution from potential
\Pgt decays)~\cite{PDG2018}.

The single top quark production processes, where a top quark is
produced with jets in the $s$ channel, $t$ channel, or $\PQt\PW$
channel, can also mimic the signal topology. The $s$-channel single
top production samples are generated using
\MGvATNLO~\cite{Alwall:2014hca} at NLO, while the $t$-channel and
$\PQt\PW$-channel samples are generated using
\POWHEG~\cite{Alioli:2009je,Re:2010bp} at NLO. The production of \PW
and \PZ bosons with jets, and vector boson pair production, are also
considered as background processes. The inclusive $\PW + \text{jets}$
and $\PZ/\PGg + \text{jets}$ samples are generated at LO using
\MGvATNLO with up to four partons included in the matrix element
calculations. The MLM technique~\cite{Alwall:2007fs} is used to avoid
the double counting of jets from the matrix element calculation and
the parton shower. The vector boson pair production samples
($\PW\PW/\PW\PZ/\PZ\PZ$, collectively referred to as ``$\PV\PV$'') are
generated using \PYTHIA at LO.

Furthermore, SM events containing only jets produced through the
strong interaction, referred to as quantum chromodynamics (QCD)
multijet events, can also produce a final state identical to the
signal topology, even though these events contain only quarks and 
gluons at the parton level. QCD multijet events can have reconstructed 
leptons from, for example, jets misidentified as isolated leptons or 
decays of bottom and charm hadrons, and \ptmiss due to the 
mismeasurement of hadronic activity inside the CMS detector.

The expected yield for each background process is determined from
simulation, with the exception of the QCD multijet background, which
is estimated from data, as described in Section~\ref{s:secEvtSel}.

\section{Object reconstruction}
\label{s:secReco}
The physics objects of interest are leptons, jets, missing
transverse momentum, vertices of $\Pp\Pp$ collisions, and displaced vertices
from the decay of bottom or charm hadrons. The particle-flow (PF)
algorithm~\cite{Sirunyan:2017ulk} is used to reconstruct these objects
by optimally using various subsystems of the CMS detector.

The collision vertices are obtained using reconstructed tracks in the
silicon tracker~\cite{Chatrchyan:2014fea}. First, candidate vertices
are obtained by clustering tracks using the deterministic annealing
algorithm. Subsequently, candidate vertices with at least two tracks
are fitted using the adaptive vertex fitter. A primary vertex
associated with a hard interaction is expected to be accompanied by a
large number of tracks. The reconstructed vertex with the largest value 
of summed physics-object $\pt^2$ is taken to be the primary $\Pp\Pp$ 
interaction vertex. The physics objects are the jets, clustered using 
the jet finding algorithm~\cite{Cacciari:2008gp,Cacciari:2011ma} with 
the tracks assigned to the vertex as inputs, and the missing transverse
momentum associated with those jets, taken as the negative vector sum
of their \pt. Further,  the reconstructed primary vertex is required to
be within 24\unit{cm} along the beam axis and within 2\unit{cm} in the
transverse direction from the nominal $\Pp\Pp$ interaction region. 

Muons, being minimum ionizing particles, can traverse a long distance
in the CMS detector. The trajectory of the muon is bent due to the
presence of a strong magnetic field inside the solenoid and the return
magnetic field in the opposite direction outside the solenoid. Muon
candidates are identified in the muon detectors and matched to tracks
measured in the silicon tracker, resulting in an excellent \pt
resolution between 1 and 10\% for \pt values up to 1\TeV~\cite{Sirunyan:2018fpa}.

Electrons are reconstructed from the tracks in the tracker and energy
deposits in the ECAL~\cite{Khachatryan:2015hwa}.
The reconstructed trajectory in the tracker is mapped to the energy
deposit in the ECAL to form an electron candidate. The bending
direction of the trajectory in the tracker is used to identify the
charge of an electron.

Because of color confinement~\cite{Polyakov:1976fu}, the quarks and
gluons produced in $\Pp\Pp$ collisions cannot exist in free states; instead,
they produce a cluster of colorless hadrons, most of which
subsequently decay to leptons and photons. As mentioned above, jets
are clustered from the PF candidates using the anti-\kt
algorithm~\cite{Cacciari:2008gp,Cacciari:2011ma} with a distance
parameter of $\Delta R = \sqrt{\smash[b]{(\Delta\eta)^2 + (\Delta\phi)^2}} = 0.4$,
where $\phi$ is the azimuthal angle.
Each jet is required to pass dedicated quality criteria to suppress
the impact of instrumental noise and misreconstruction. Additional $\Pp\Pp$
interactions within the
same or nearby bunch crossings (pileup) can contribute extra tracks
and calorimetric energy deposits, increasing the apparent jet
momentum. To mitigate this effect, tracks identified to be originating
from pileup vertices are discarded and an offset correction is applied
to correct for remaining contributions~\cite{Sirunyan:2017ulk}. Jet
energy corrections are derived from simulation studies so that the
average measured response of jets becomes identical to that of
particle-level jets. In situ measurements of the momentum balance in
dijet, $\PGg+\text{jet}$, $\PZ+\text{jet}$, and multijet events are
used to determine any residual differences between the jet energy
scale in data and in simulation, and appropriate corrections are
applied~\cite{Khachatryan:2016kdb}.

The missing transverse momentum vector \ptvecmiss is defined as the
projection onto the plane perpendicular to the beam axis of
the negative vector sum of the momenta of all PF objects in an event.
Its magnitude is referred to as \ptmiss. Neutrinos, being weakly
interacting particles with a very low cross section, cannot be
directly detected by the CMS detector and thus contribute to \ptmiss.
The reconstruction of \ptmiss is improved by propagating the jet
energy corrections to it.

There are two \PQb jets in the final state as illustrated in
Fig.~\ref{fig:feyn_diag_sig}, in both the charged Higgs boson signal
process and the SM \ttbar background. An accurate
identification of \PQb jets substantially reduces the SM backgrounds
from other processes, such as $\PZ/\PGg + \text{jets}$, $\PV\PV$,
or $\PW + \text{jets}$. The combined secondary vertex (CSV)
algorithm~\cite{Sirunyan:2017ezt} is used to tag a \PQb jet. The
algorithm combines information on track impact parameters and
secondary vertices within a jet into an artificial neural network
classifier that provides separation between a \PQb jet and jets of
other flavors. As the charged Higgs boson decays to a charm and a
strange antiquark, the identification of charm jets is expected to
increase the signal significance. A charm tagger has been
developed~\cite{Sirunyan:2017ezt}, which is based on the CSV method
and works similarly to the \PQb tagging procedure.

The \pt of jets in the simulated samples is corrected 
using the jet energy scale (JES) and jet energy resolution (JER) 
data-to-simulation scale factors~\cite{Khachatryan:2016kdb}. The 
lepton reconstruction, \PQb, and \PQc tagging efficiencies are 
different in data and simulated samples; to correct for this, the 
corresponding data-to-simulation scale factors are applied to the 
simulated events.

\section{Event selection}
\label{s:secEvtSel}
In the event topology of interest, there are four jets (two \PQb jets,
one \PQc jet, and one light-flavor jet), one charged lepton, and
\ptmiss. Various selection requirements are applied to ensure the
resulting events have this topology.

The online event selection requires, at the L1 trigger level, either a
muon candidate with $\pt > 22\GeV$ or electron/photon candidate with
$\pt > 30\GeV$ (22\GeV if it is isolated); at the HLT level, an
isolated muon (electron) with $\pt > 24$ (27)\GeV is required.
The relative isolation ($I_\text{rel}$) of a lepton is defined as the
ratio of the sum of $\pt$ for all the other particles within a cone of
$\Delta R = 0.4$ around the lepton direction, divided by the lepton
$\pt$ after correcting for the contribution from
pileup~\cite{Sirunyan:2018fpa,Cacciari:2007fd}.

In the offline analysis, events that pass the trigger selection and
contain a muon (electron) with $\pt > 26$ (30)\GeV and $\abs{\eta} < 2.4$
(2.5) are selected. To eliminate events where the lepton is found
within a jet, the muon is required to have $I_\text{rel}^{\Pgm} < 0.15$
and the electron is required to have $I_\text{rel}^{\Pe} < 0.08$
(0.07) in the barrel (endcap) regions. No charge requirement is
applied to the lepton. The signal event topology has only one lepton,
so events having a second muon with $\pt^{\PGm} > 15\GeV$, $\abs{\eta} < 2.4$,
and $I_\text{rel}^{\Pgm} < 0.25$, or an electron with $\pt^{\Pe} > 15\GeV$,
$\abs{\eta} < 2.5$, and $I_\text{rel}^{\Pe} < 0.18$ (0.16) in the barrel
(endcap) regions, are rejected.

Jets are selected by requiring $\pt^\text{j} > 25\GeV$,
$\abs{\eta^\text{j}} < 2.4$, neutral hadron energy fraction $<$ 0.99,
neutral electromagnetic energy fraction $<$ 0.99, number of
constituents $>$ 1, charged hadron energy fraction $>$ 0,
charged-hadron multiplicity $>$ 0, and charged-hadron electromagnetic
energy fraction $<$ 0.99, as detailed in Ref.~\cite{Sirunyan:2017ulk};
at least four jets are required. The \ptmiss must exceed 20\GeV. The
events are required to have at least two \PQb jets with a selection that
has 63\% \PQb tagging efficiency~\cite{Sirunyan:2017ezt}. The corresponding 
probability of a light-flavor (charm) jet being misidentified as a \PQb 
jet is 1 (12)\%, where ``light flavor'' refers to jets originating from 
\PQu, \PQd, \PQs, or \Pg. The events are categorized depending on the 
charm tagging results for the jets, as discussed in Section~\ref{s:secKF}.

To estimate QCD multijet background, a matrix method 
based on the two uncorrelated variables $I_\text{rel}$ 
and \ptmiss, also known as an ``ABCD'' method, is used, which proceeds 
as follows. First, a normalization is determined from the (low \ptmiss, isolated) and
(low \ptmiss, anti-isolated) regions; then the QCD background
distribution is determined from the (high \ptmiss, anti-isolated)
region. By using the normalization obtained on the distribution,
the expected QCD multijet contribution is determined in the signal
region (high \ptmiss, isolated). The low- and high-\ptmiss regions are
defined by $\ptmiss < 20\GeV$ and $\ptmiss > 20\GeV$, respectively. In
the muon channel, the isolated and anti-isolated regions are defined by
$I_\text{rel}^{\Pgm} < 0.15$ and $0.15 < I_\text{rel}^{\Pgm} < 0.4$, respectively. For the
electron channel, the isolated region corresponds to
$I_\text{rel}^{\Pe} < 0.08$ (0.07) and the anti-isolated region to
0.08 (0.07) $< I_\text{rel}^{\Pe} < 0.3$ for electrons in the barrel
(endcap) regions. The QCD multijet background is 
estimated after applying both \PQb and \PQc tagging.

\section{Dijet invariant mass distribution}
\label{s:secKF}
\begin{figure*}
    \centering
    \includegraphics[width=0.49\textwidth]{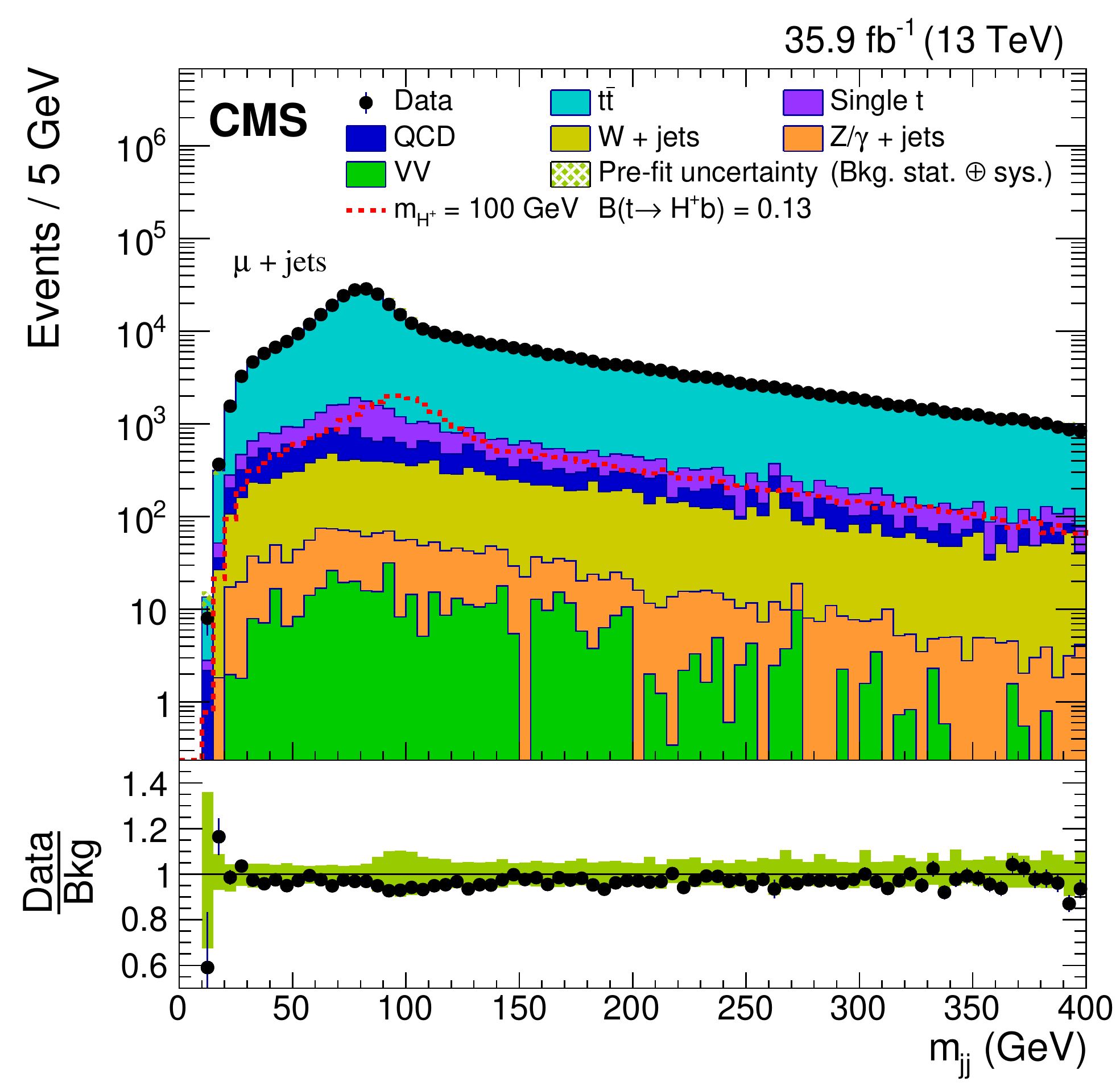}
    \includegraphics[width=0.49\textwidth]{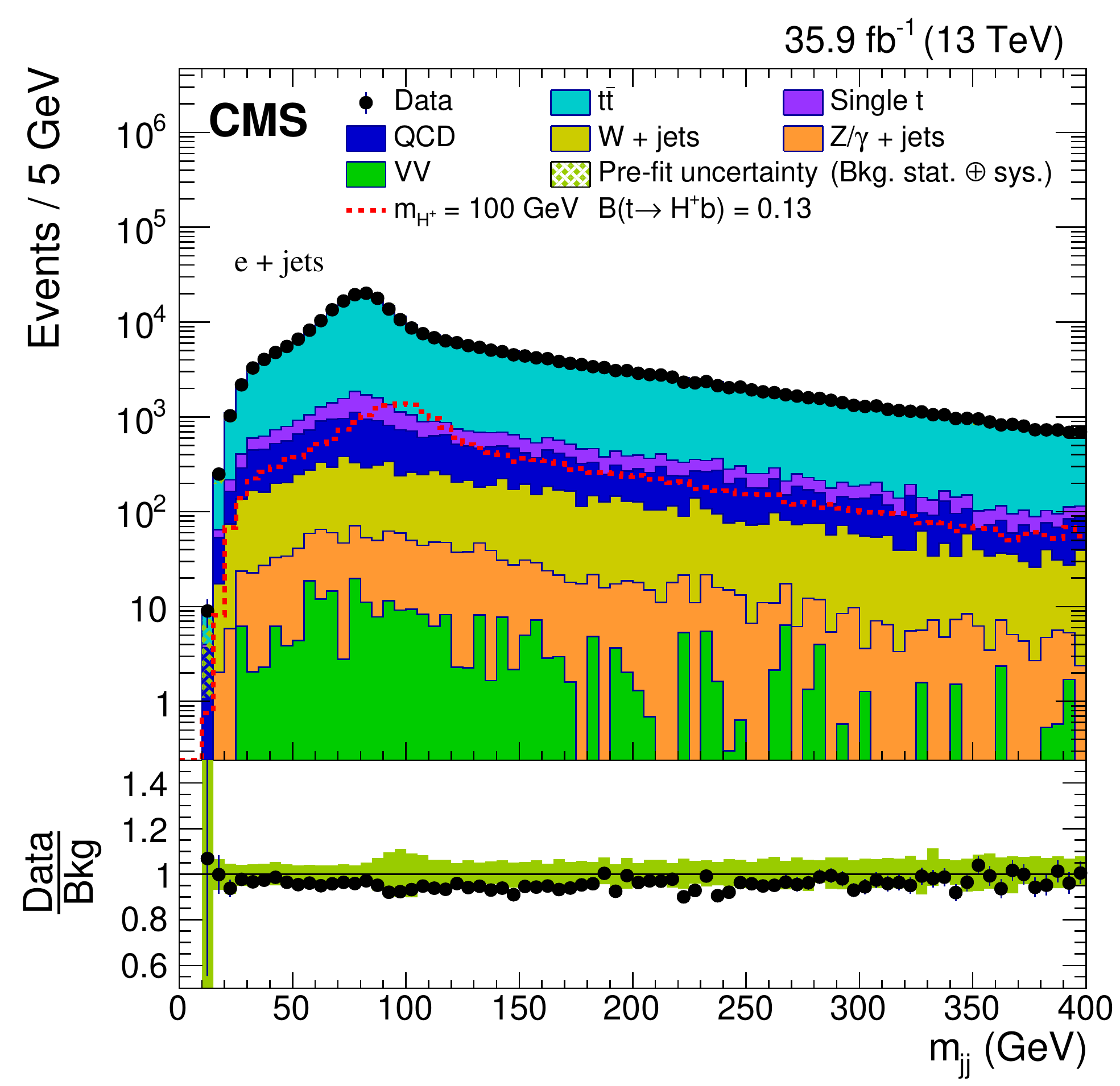}
    \includegraphics[width=0.49\textwidth]{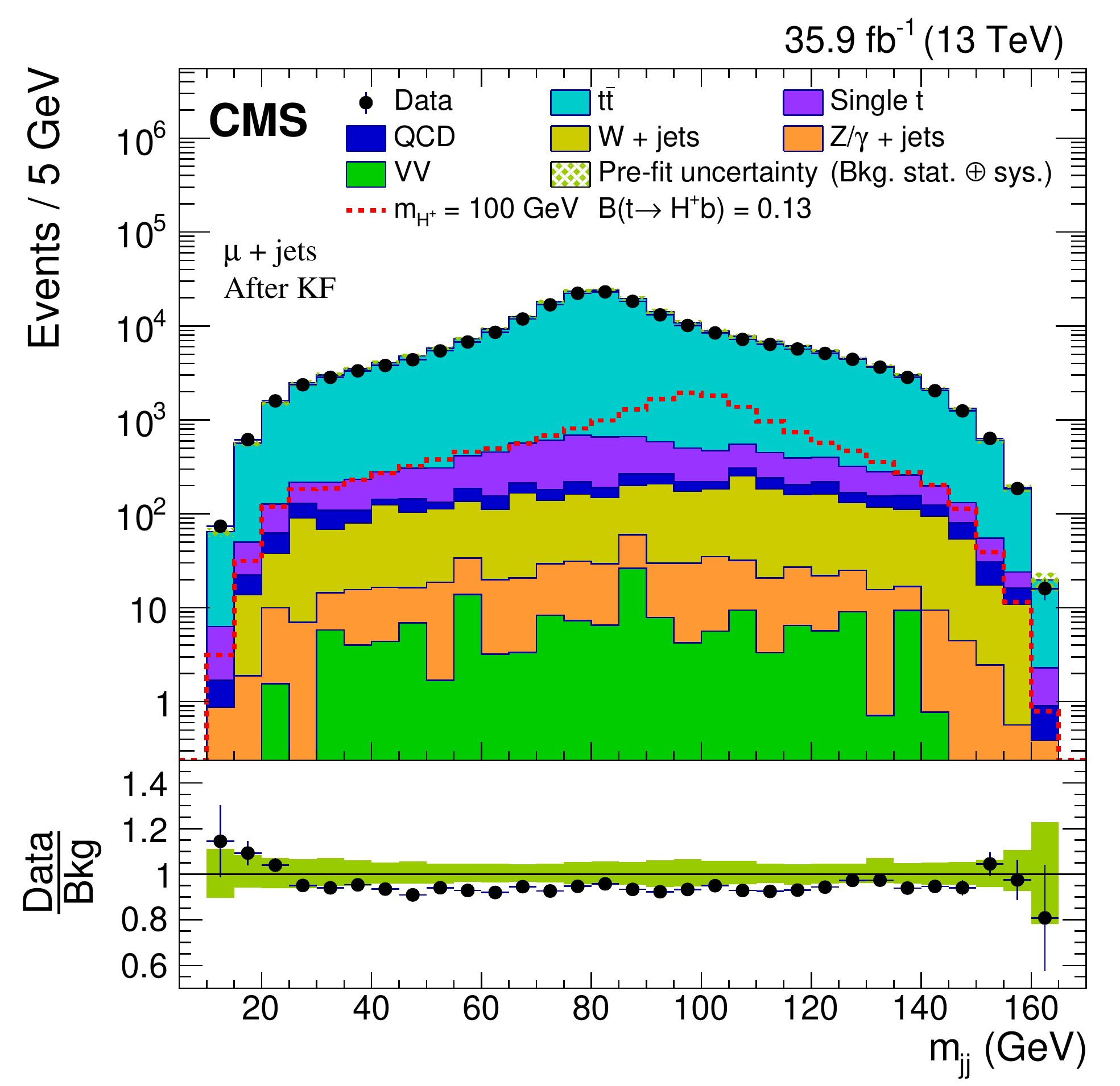}
    \includegraphics[width=0.49\textwidth]{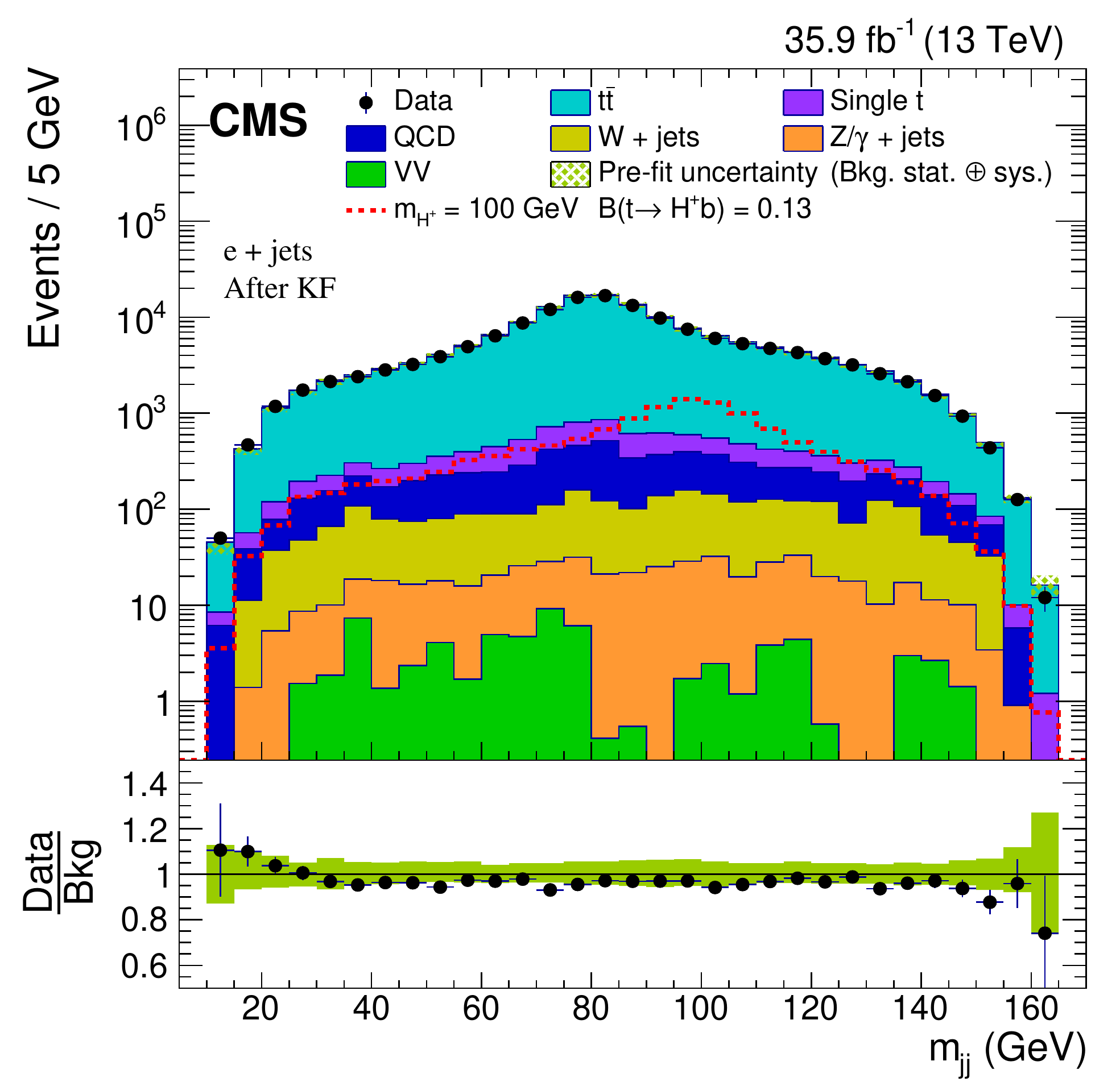}
    \caption{Distributions of \mjj, prior to the fit to data, of the
        two highest \pt non-\PQb jets for the muon\,+\,jets channel
        (left column) and the electron\,+\,jets channel (right column).
        The two distributions in the upper row are obtained using
        reconstructed jets. The distributions in the lower row are
        calculated using jets after the kinematic fit. 
        The uncertainty band (showing the absolute
        uncertainty in the upper panels, and the relative uncertainty in
        the lower panels) includes both statistical and systematic
        components. The signal events are scaled by twice the maximum
        observed upper limit on \Bthb
        obtained at 8\TeV~\cite{Khachatryan:2015uua}.}
    \label{fig:mjjBTagKinFit}
\end{figure*}

The invariant mass of the system of the two non-\PQb jets (\mjj),
assumed to be $\PQc\PAQs$ or $\PAQc\PQs$, is used as the final
observable. The \mjj distribution of the two highest-\pt non-\PQb
jets is shown in the top row of Fig.~\ref{fig:mjjBTagKinFit} for the
two leptonic channels. If the two observed non-\PQb jets come from
a semileptonic \ttbar decay, then the \mjj distribution should have a
peak at the \PW boson mass. The observed mean of the \mjj distribution
is much higher (around 138\GeV), reflecting the fact that the two
non-\PQb jets in each event may not necessarily come from the decay of a
\PW boson.

To identify semileptonic \ttbar events, a kinematic fit (KF) is
performed on the reconstructed objects using the top quark kinematic
fitter package~\cite{DHondt:926540}. The top kinematic fitter takes
physics objects such as leptons, jets, \ptmiss, and their resolutions
as input, and gives improved four-vectors of leptons, jets, and a
neutrino, along with the overall $\chi^2$ and fit probability for the
event, as the output. The $x$ and $y$ components of the neutrino
momentum are taken from \ptmiss, as the missing transverse momentum
is attributed to the neutrino, and the $z$ component of the neutrino
momentum, $p_z^{\PGn}$, is determined from the fit. The following
kinematic constraints are imposed on the semileptonic \ttbar system:
\begin{linenomath}
\begin{subequations}
\label{eq:constraintKF}
\begin{align}
	m_{\text{inv}}(\PQb_{\text{had}}\PQq\PAQq) = m_{\PQt} = 172.5\GeV
    \label{eq:constraintKF1}\\
	m_{\text{inv}}(\PQb_{\text{lep}}\ell\PGn_\ell) = m_{\PAQt} = 172.5\GeV ,
    \label{eq:constraintKF2}
\end{align}
\end{subequations}
\end{linenomath}
\noindent where $m_{\text{inv}}$ is the corresponding invariant mass
and $\PQb_{\text{had (lep)}}$ is the \PQb quark produced by the
hadronic (leptonic) top decay. After the fit, $p_z^{\PGn}$ is
determined from Eq.~(\ref{eq:constraintKF2}). For every event, a
$\chi^2$ is constructed and minimized by varying the \pt, $\eta$, and
$\phi$ of each object within their resolution. The values of \pt,
$\eta$, and $\phi$ are finally selected that minimize the $\chi^2$ and
at the same time satisfy Eq.~(\ref{eq:constraintKF}). In the output,
the top quark kinematic fitter gives exactly four jets (two \PQb jets,
one from each of the leptonic and hadronic \PQt decays, and two
non-\PQb jets from the hadronic \PQt decay), a lepton, and a
neutrino. No cut is placed on $\chi^2$ and events for which the fit
does not converge are discarded.

Also, the same kinematic requirements (on $\pt$, $\eta$, and $I_\text{rel}$) 
as for the reconstructed objects are applied to the fitted objects. 
The directions of the kinematically fitted jets and lepton are required 
to be compatible with those of the reconstructed jets and lepton ($\Delta R < 0.2$),
respectively. The efficiency of the KF selection for data, simulated \ttbar, and
simulated signal events is 43, 47, and 49\%, respectively. 
The \mjj distributions after the KF selection are shown in the bottom row
of Fig.~\ref{fig:mjjBTagKinFit}, showing that the mean of the \mjj 
distribution is closer to the \PW boson mass.

\begin{figure*}
\centering
{\includegraphics[width=0.4\textwidth]{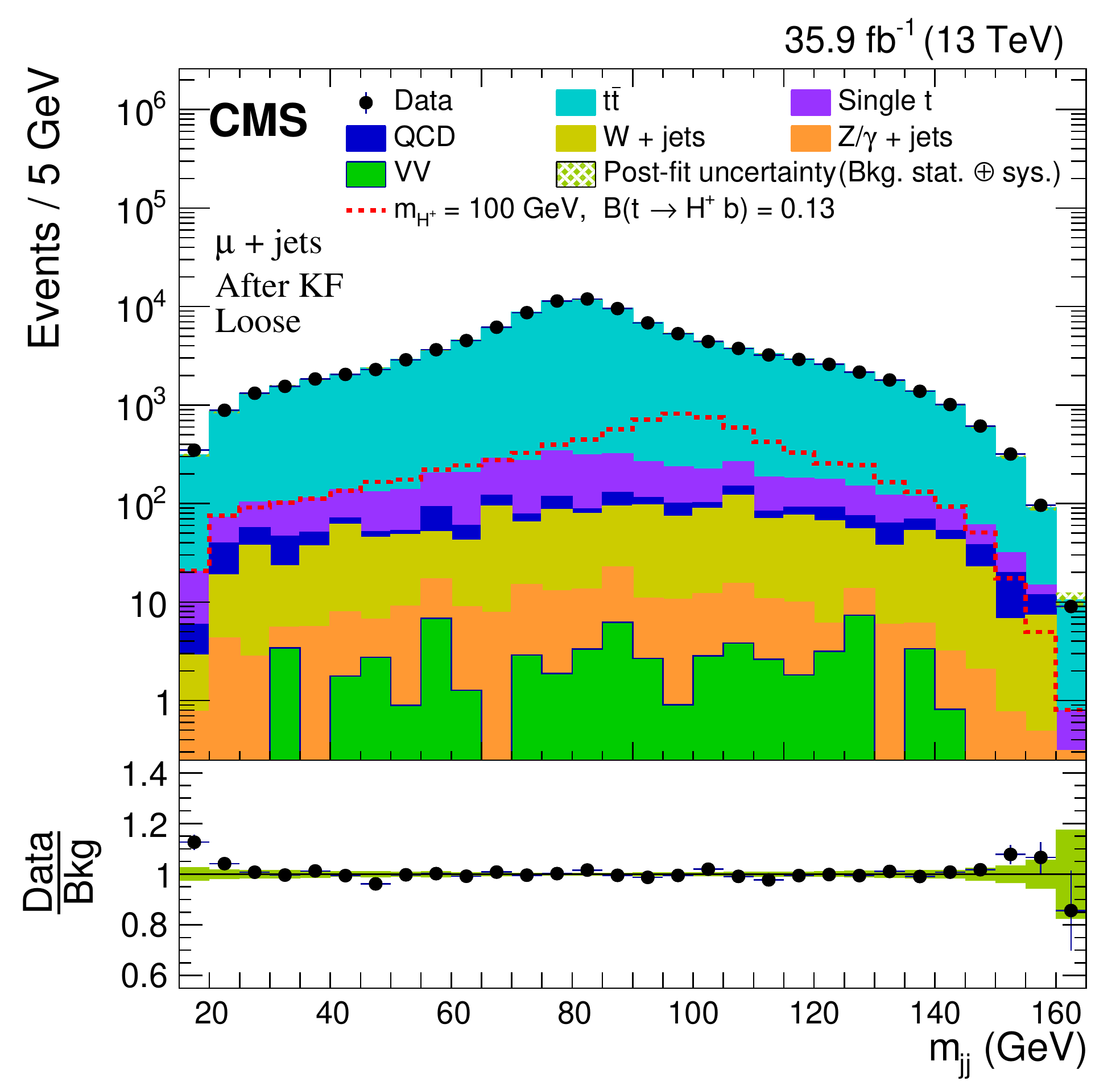}}
{\includegraphics[width=0.4\textwidth]{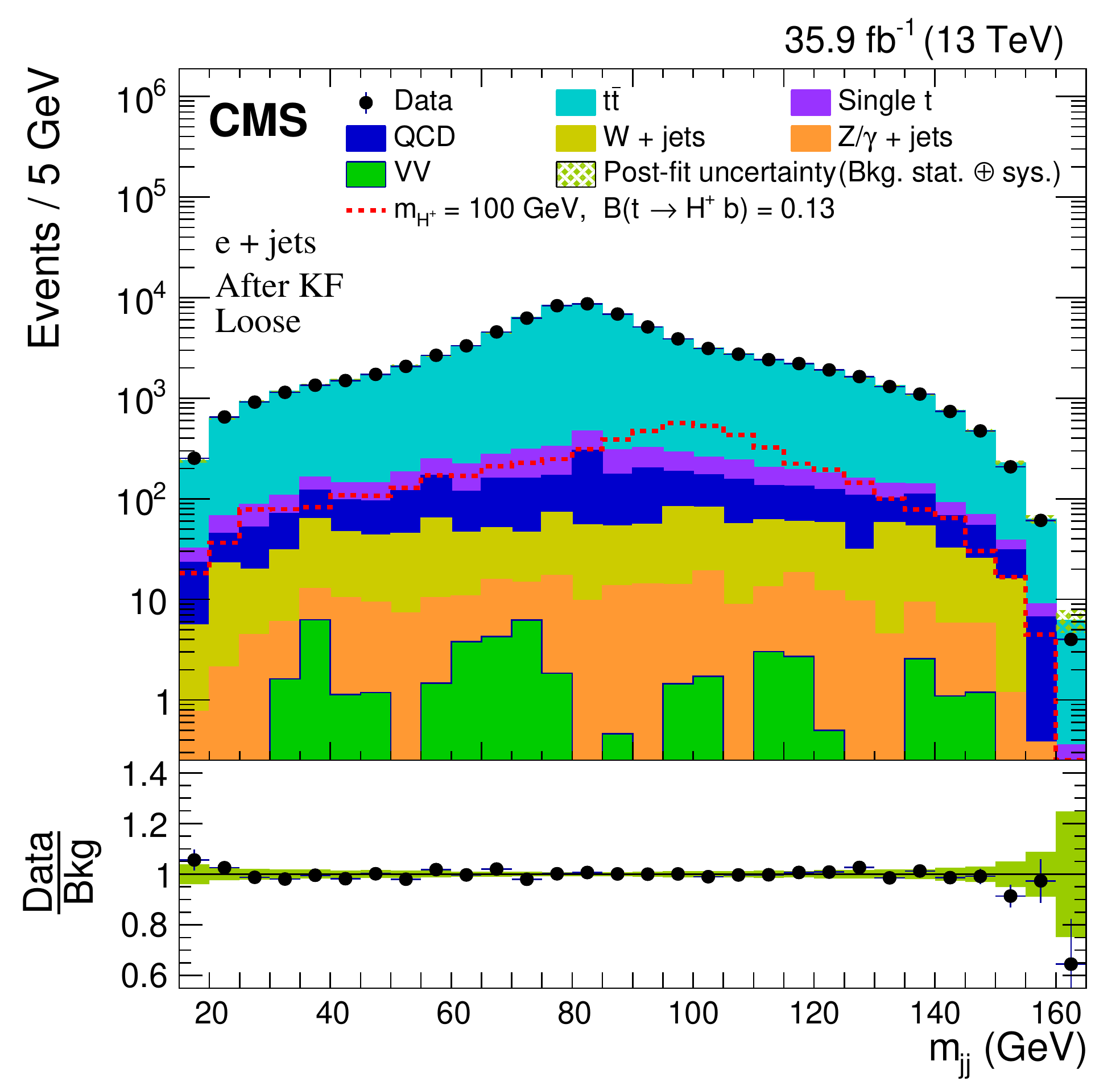}}
{\includegraphics[width=0.4\textwidth]{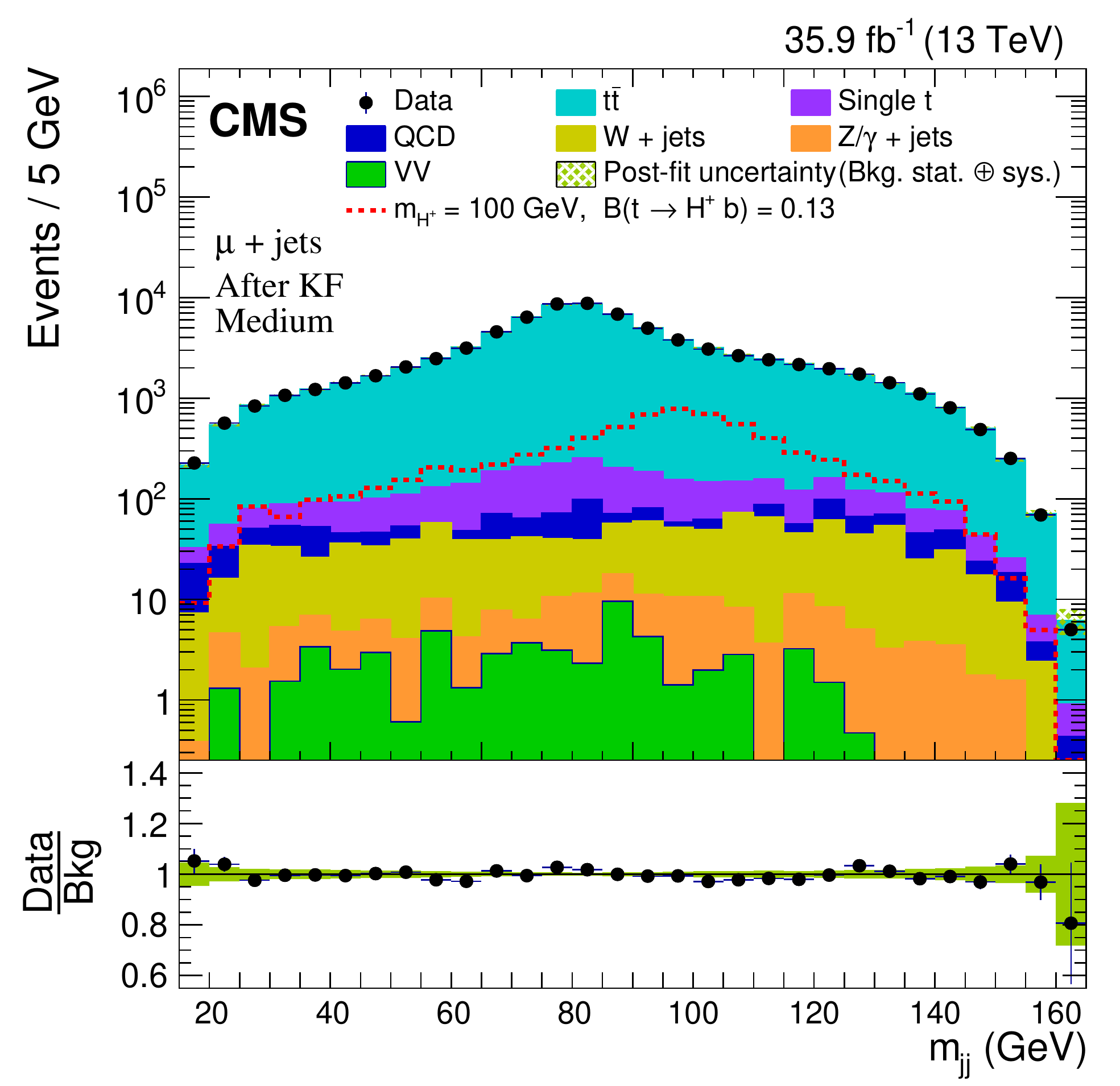}}
{\includegraphics[width=0.4\textwidth]{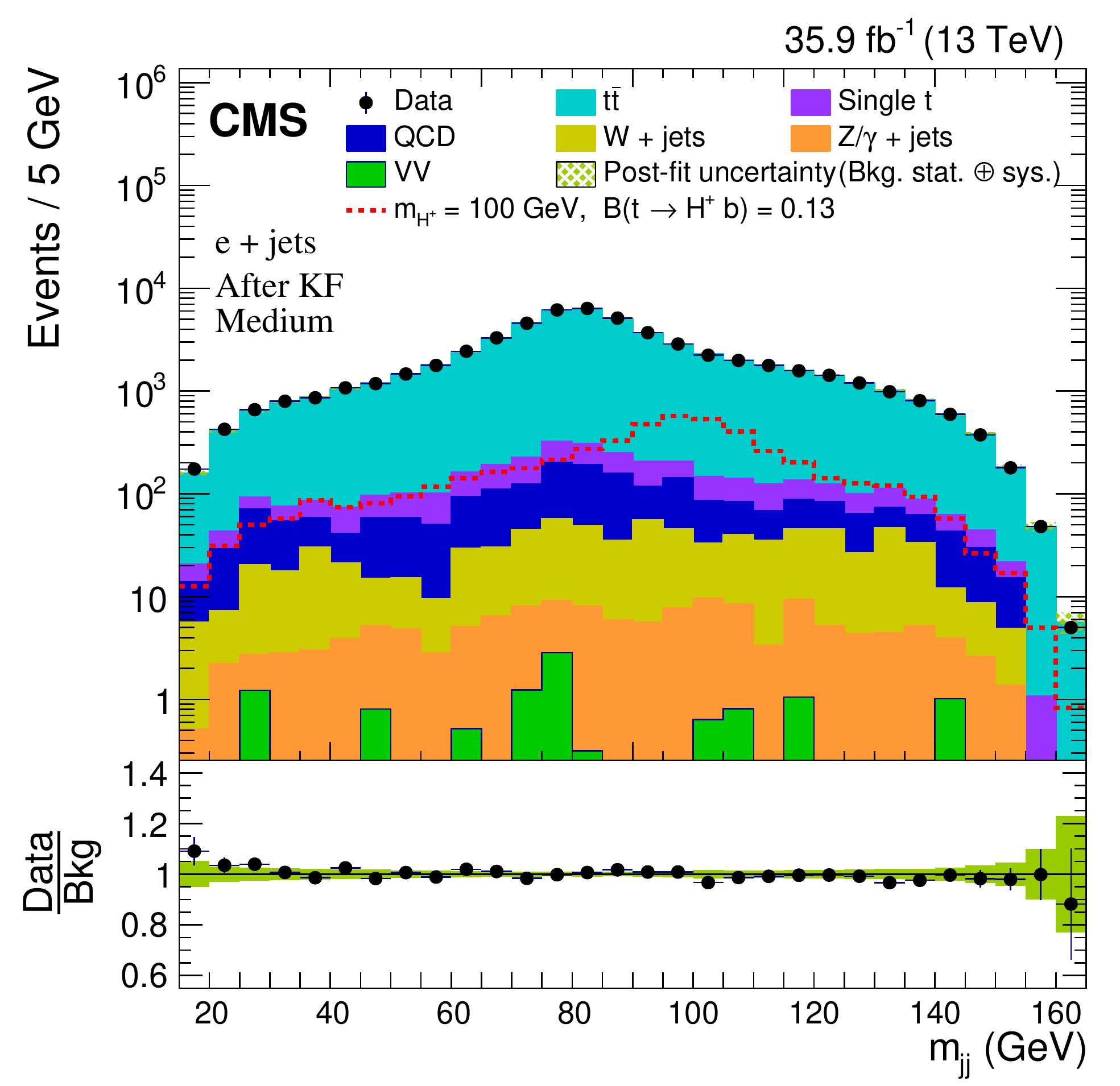}}
{\includegraphics[width=0.4\textwidth]{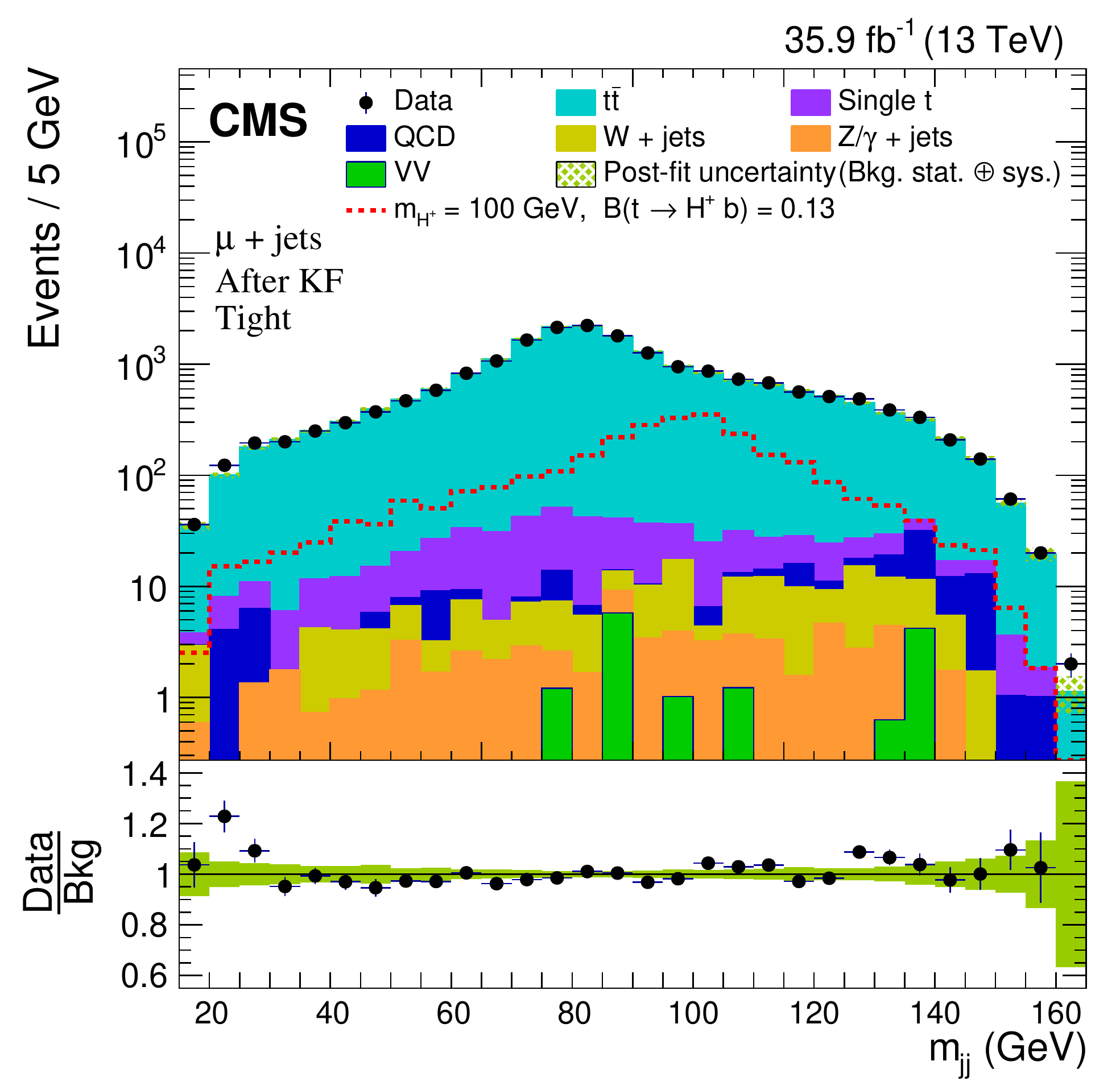}}
{\includegraphics[width=0.4\textwidth]{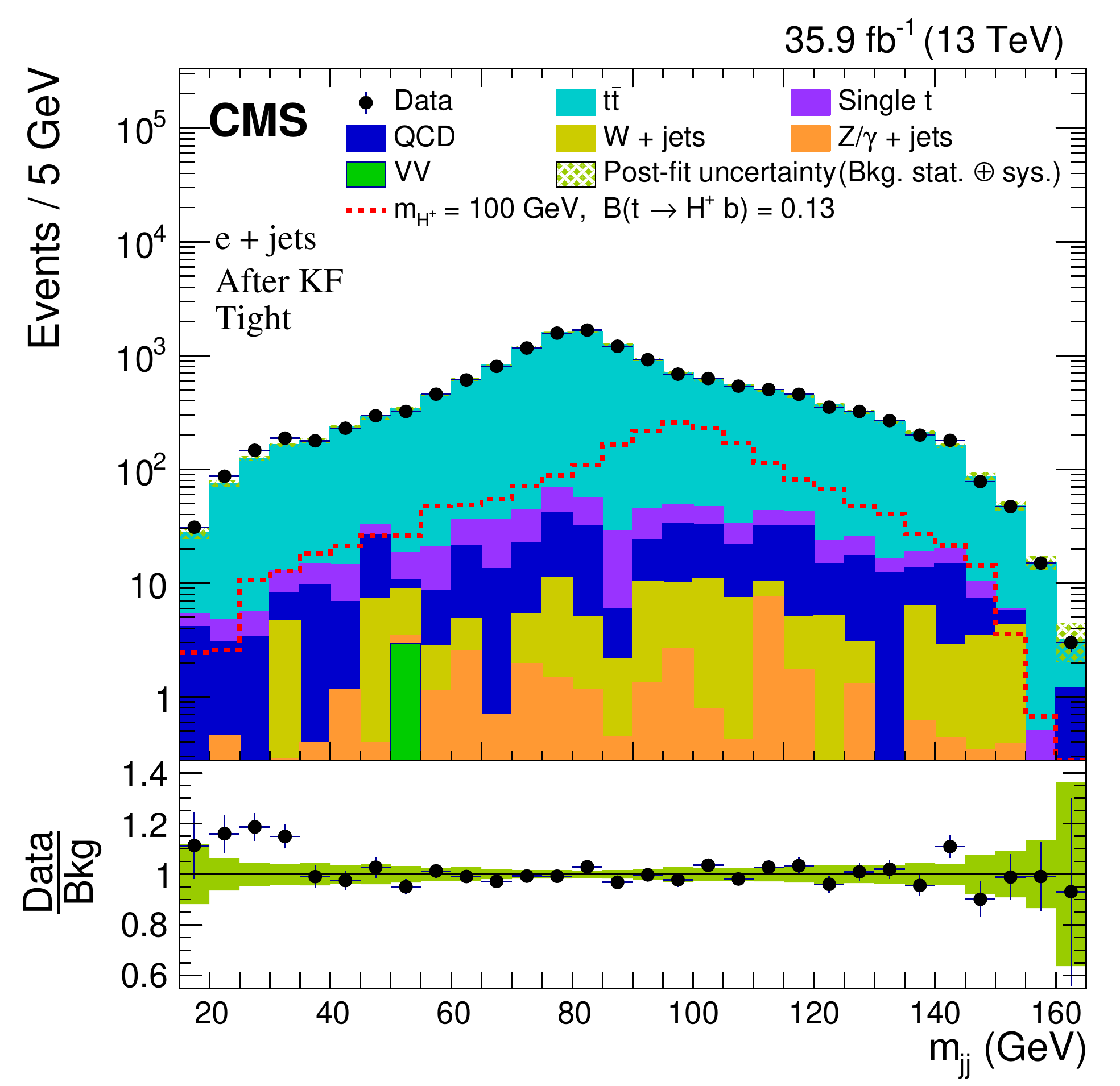}}
\caption{Distributions of \mjj, after a background-only fit to the
    data, in the exclusive charm tagging categories for the muon\,+\,jets
    (left column) and electron\,+\,jets (right column) channels.
    The upper row shows the exclusive loose category, the middle row
    shows the exclusive medium category, and the lower row shows the
    exclusive tight category. The expected signal significance (prior
    to the fit) can be observed to vary across the different
    categories. The uncertainty band (showing the absolute
    uncertainty in the upper panels, and the relative uncertainty in
    the lower panels)
    includes both statistical and systematic components after
    the background-only fit. The signal distributions are scaled by twice the
    maximum observed upper limit on \Bthb
    obtained at 8\TeV~\cite{Khachatryan:2015uua}.}
\label{fig:mjj_cTagEx}
\end{figure*}

The two non-\PQb jets coming from the hadronic \PQt decay are
further used for charm tagging. There are three \PQc tagging working
points (loose, medium, and tight) based on the efficiency
of a \PQc quark being tagged as a \PQc jet~\cite{Sirunyan:2017ezt}.
The corresponding efficiencies are shown in Table~\ref{tab:cTagEff}.
\begin{table}
\centering
\topcaption{The efficiency of the \PQc jet tagger to tag a jet from a \PQc quark
($\epsilon^{\PQc}$), a \PQb quark ($\epsilon^{\PQb}$), or light flavor
($\epsilon^{\PQu\PQd\PQs\Pg}$) at different working points, as determined from
simulation~\cite{Sirunyan:2017ezt}.}
\label{tab:cTagEff}
\begin{scotch}{cccccc}
Working point & $\epsilon^{\PQc}$ (\%) & $\epsilon^{\PQb}$ (\%) &
$\epsilon^{\PQu\PQd\PQs\Pg}$ (\%) \\ \hline
Loose  & 88 & 36 & 91 \\
Medium & 40 & 17 & 19 \\
Tight  & 19 & 20 & 1.2\\
\end{scotch}
\end{table}
The events are divided exclusively into loose, medium, and tight
categories, based on whether at least one of the non-\PQb jets
passes the loose but neither passes the medium, at least one passes
the medium but neither passes the tight, or at least one passes the
tight working points of the charm tagging selection requirements shown
in Table~\ref{tab:cTagEff}, respectively. The \mjj distributions for
the exclusive charm categories are shown in Fig.~\ref{fig:mjj_cTagEx}
after a background-only maximum likelihood fit to data. From these
figures, it can be seen that the expected signal-to-background ratio 
increases for the charm categories with tighter requirements,
so partitioning the events into categories results in an enhanced 
signal sensitivity. Table~\ref{tab:eventYield} shows the corresponding
event yields for the different  charm categories after the background-only
fit to the data reported in Section~\ref{s:secResults}, with statistical and systematic 
uncertainties as discussed in Section~\ref{s:secSys}.

\begin{table*}[htb]
  \centering
\topcaption{Expected event yields for different signal mass scenarios and
    backgrounds in each of the channels and event categories. The
    number of events is shown along with its uncertainty, including
    statistical and systematic effects. The yields of the
    background processes are obtained after a background-only fit to
    the data. The total uncertainty in the background process is
    calculated by taking into account all the positive as well as
    negative correlations among the fit parameters. The signal event yields
    are scaled by twice the maximum observed upper limit on
    \Bthb obtained at
    8\TeV~\cite{Khachatryan:2015uua}.}
\label{tab:eventYield}
\renewcommand{\arraystretch}{1.2}
  \cmsTable{
\begin{scotch}{cccccccc}
\multicolumn{1}{c}{Process} & \multicolumn{2}{c}{Loose} & \multicolumn{2}{c}{Medium} & \multicolumn{2}{c}{Tight} \\
                  & \mujets   &  \ejets      & \mujets   &  \ejets      & \mujets   &  \ejets \\
\hline
$\mhp=80\GeV$ & $7690 \pm 550$ & $5430 \pm 380$ & $6560 \pm 490$ & $4700 \pm 370$ & $2670 \pm 270$ & $1860 \pm 180$\\
$\mhp=90\GeV$ & $7710 \pm 550$ & $5620 \pm 400$ & $6770 \pm 510$ & $4860 \pm 380$ & $2630 \pm 260$ & $1870 \pm 190$\\
$\mhp=100\GeV$ & $7950 \pm 590$ & $5550 \pm 400$ & $7070 \pm 540$ & $4950 \pm 360$ & $2770 \pm 270$ & $2000 \pm 200$\\
$\mhp=120\GeV$ & $7620 \pm 570$ & $5360 \pm 400$ & $6870 \pm 510$ & $4780 \pm 360$ & $2650 \pm 260$ & $1960 \pm 190$\\
$\mhp=140\GeV$ & $6160 \pm 500$ & $4370 \pm 360$ & $5420 \pm 420$ & $3840 \pm 310$ & $2010 \pm 210$ & $1500 \pm 150$\\
$\mhp=150\GeV$ & $4530 \pm 390$ & $3230 \pm 280$ & $3850 \pm 330$ & $2800 \pm 250$ & $1340 \pm 140$ & $1030 \pm 120$\\
$\mhp=155\GeV$ & $3700 \pm 340$ & $2560 \pm 250$ & $2980 \pm 270$ & $2230 \pm 220$ & $1020 \pm 120$ & $766 \pm 86$\\
$\mhp=160\GeV$ & $2780 \pm 270$ & $2080 \pm 200$ & $2370 \pm 230$ & $1710 \pm 180$ & $728 \pm 83$ & $510 \pm 59$\\
[\cmsTabSkip]
\ttbar & $100540 \pm 410$ & $71800 \pm 470$ & $73210 \pm 320$ & $52340 \pm 290$ & $18760 \pm 130$ & $13380 \pm 130$ \\
Single \PQt quark & $2750 \pm 220$ & $1970 \pm 160$ & $1940 \pm 160$ & $1400 \pm 110$ & $421 \pm 35$ & $302 \pm 26$ \\
QCD multijet & $520 \pm 130$ & $2120 \pm 470$ & $498 \pm 98$ & $1460 \pm 210$ & $88 \pm 28$ & $346 \pm 39$ \\
$\PW + \text{jets}$ & $1360 \pm 140$ & $1061 \pm 90$ & $950 \pm 110$ & $681 \pm 58$ & $127 \pm 23$ & $102 \pm 9$ \\
$\PZ/\PGg +\text{jets}$ & $189 \pm 18$ & $240 \pm 25$ & $132 \pm 13$ & $132 \pm 14$ & $56 \pm 7$ & $31 \pm 4$ \\
$\PV\PV$ & $61 \pm 9$ & $43 \pm 6$ & $56 \pm 8$ & $11 \pm 4$ & $15 \pm 5$ & $3 \pm 1$ \\
[\cmsTabSkip]
All background & $105410 \pm 500$ & $77240 \pm 690$ & $76780 \pm 390$ & $56020 \pm 380$ & $19470 \pm 140$ & $14160 \pm 140$ \\
[\cmsTabSkip]
Data & 105474 & 77244 & 76807 & 56051 & 19437 & 14179 \\
\end{scotch}
}
\end{table*}

\section{Systematic uncertainties}
\label{s:secSys}
There are various sources of systematic uncertainty, which may arise
due to detector calibration effects, uncertainty in the measured
reconstruction efficiency, the theoretical modeling of signal events,
and other effects.

The uncertainty in the integrated luminosity is 2.5\%~\cite{CMS-PAS-LUM-17-001}. Each distribution for
simulated events is normalized to the expected number of events in data, using the factor
$L_\text{data}\sigma_\text{sim}/N_\text{sim}$, where $L_\text{data}$ is the integrated luminosity of the data
sample, $N_\text{sim}$ is the total number of events in the simulated sample, and $\sigma_\text{sim}$ is the
cross section for the simulated process considered; the uncertainties in $\sigma_\text{sim}$ thus contribute
to the uncertainty in each background prediction. The uncertainties in 
$\sigma_\text{sim}$ for \ttbar, single \PQt quark, $\PW + \text{jets}$, 
$\PZ/\PGg +\text{jets}$, and $\PV\PV$ processes are 6.1, 7.0, 4.5, 
5.0, and 4.0\%, respectively. To account for the uncertainty in the 
pileup distribution, the total inelastic cross section of 69.2\unit{mb} 
is varied by its uncertainty of 4.7\%~\cite{Sirunyan:2018nqx} and the 
simulated events are reweighted to match the pileup distribution in 
the data. The systematic uncertainty in the data-to-simulation scale 
factor for the lepton reconstruction efficiencies is 3.0\% for both 
muons and electrons~\cite{Sirunyan:2018fpa,Khachatryan:2015hwa}. 

The systematic uncertainties due to JES and JER data-to-simulation 
scale factors in the \pt of the jets and \ptmiss are estimated by 
varying these within their uncertainties~\cite{Khachatryan:2016kdb}. 
The \PQb and \PQc tag data-to-simulation scale factors are varied 
within their uncertainties to estimate the corresponding uncertainties, 
with correlations applied~\cite{Sirunyan:2017ezt}. 

To estimate the systematic uncertainty in the QCD multijet background
estimation, the muon (electron) relative isolation threshold is
conservatively changed to 0.17 (0.11) and the corresponding changes in
the QCD yields are determined.

It is found that the \pt distribution of \PQt quarks in \ttbar events
in data is softer compared to that in simulated
samples~\cite{Khachatryan:2016mnb}. This is corrected by applying the
following weight as a function of \pt for SM \ttbar and charged Higgs
boson signal samples:
\ifthenelse{\boolean{cms@external}}{
\begin{linenomath}
\begin{multline}
w_\text{t}=\sqrt{{\text{SF}(\PQt)\text{SF}(\PAQt)}},\\
\text{with SF}\equiv\exp(0.09494 -0.00084\pt).
\label{eq:top_wt}
\end{multline}
\end{linenomath}
}{
\begin{linenomath}
\begin{equation}
w_\text{t}=\sqrt{{\text{SF}(\PQt)\text{SF}(\PAQt)}},\ \text{with SF}\equiv\exp(0.09494 -0.00084\pt).
\label{eq:top_wt}
\end{equation}
\end{linenomath}
}
The values in the exponent are derived in Ref.~\cite{Sirunyan:2018ucr}.
The generator-level \pt of the \PQt and \PAQt are used to calculate
SF. To evaluate the systematic uncertainty due to $w_\text{t}$, it is
varied to 1 and $w_\text{t}^2$.

The SM \ttbar sample was generated with $\mt = 172.5\GeV$. To evaluate
the effect of the chosen \mt on the \mjj distribution, alternate
\ttbar samples with $\mt = 171.5$ and 173.5\GeV are considered. To
observe the effect of NLO matrix element parton shower matching,
additional SM \ttbar samples are generated by changing the default
damping parameter $h_{\text{damp}}$ value of 1.58\mt to
2.24\mt and \mt~\cite{CMS-PAS-TOP-16-021}.
Similarly, SM \ttbar samples where the common nominal value of
renormalization and factorization scales is simultaneously changed by
factors of 0.5 and 2 are used to evaluate the uncertainties due to
these scales~\cite{Butterworth:2015oua}. The systematic uncertainties
due to \PQt quark mass, parton shower matching, and renormalization
and factorization scales are in the ranges 0.2--3.3, 0.7--1.9, and
0.4--1.6\%, respectively, depending on the channel and charm tagging
category.

The signal extraction procedure is based on a binned maximum likelihood fit of
the \mjj distributions, as described in Section~\ref{s:secKF}. The systematic
uncertainties prior to the fit on the different process yields are listed in
Table~\ref{tab:sysUnc}, when they differ from process to process. All systematic
uncertainties are incorporated into the fit as nuisance parameters, where the
effect of each systematic uncertainty on the overall normalization of the \mjj
distribution is included as a lognormal probability distribution.  The
statistical uncertainties in the total yield of all backgrounds and the signal
samples are also shown in Table~\ref{tab:sysUnc}. However, these are not
incorporated in the likelihood. To account for the statistical uncertainty in
each bin of \mjj, one nuisance parameter per bin is considered for the sum of
all backgrounds and charged Higgs boson samples~\cite{Barlow:1993dm}.

\begin{table*}[ht]
  \centering
\topcaption{Systematic and statistical uncertainties in the event yield for the
different processes in \%, when they differ from process to process, prior to
the fit to data, for the exclusive charm categories in the muon (electron)
channel. The ``\NA'' indicates that the corresponding uncertainties are either
not considered for the given process, or too small to be measured.}
\label{tab:sysUnc}
\renewcommand{\arraystretch}{1.2}
  \cmsTable{
\begin{scotch}{cccccccc}
Category &Process& Pileup & jet \& \ptmiss & \PQb \& \PQc jets & Normalization& Statistical & \pt (\PQt)\\
\hline
 Loose  & $\mhp=100\GeV$ &  0.6 (1.1) & 4.2 (3.5) &  6.1 (6.1) & 6.1 (6.1) & 1.0 (1.2) & 1.4 (1.8) \\
 & \ttbar &  0.9 (1.1) & 3.6 (3.6) &  5.8 (5.8) & 6.1 (6.1) & 0.2 (0.2) & 1.5 (1.9) \\
 & Single \PQt quark &  0.6 (0.8) & 4.9 (5.4) &  6.5 (6.6) & 5.0 (5.0) & 0.7 (0.8) & \NA \\
 & $\PW + \text{jets}$ &  2.3 (0.4) & 13 (6.9) &  10 (10) & 5.0 (5.0) & 3.9 (4.5) & \NA \\
 & $\PZ/\PGg + \text{jets}$ &  1.8 (2.4) & 11 (8.4) &  9.2 (9.0) & 4.5 (4.5) & 5.7 (4.2) & \NA \\
 & $\PV\PV$ &  1.5 (7.9) & 19 (13) &  7.2 (7.0) & 4.0 (4.0) & 19 (22) & \NA \\
 & QCD multijet &  \NA & \NA &  \NA & 10 (10) & 20 (7.3) & \NA \\
[\cmsTabSkip]
 Medium & $\mhp=100\GeV$ &  0.4 (0.3) & 3.5 (2.0) &  6.7 (6.8) & 6.1 (6.1) & 1.1 (1.3) & 1.6 (1.9) \\
 & \ttbar &  0.3 (0.4) & 3.0 (3.0) &  7.3 (7.3) & 6.1 (6.1) & 0.2 (0.3) & 1.5 (2.0) \\
 & Single \PQt quark &  0.3 (0.1) & 4.4 (4.1) &  8.1 (8.1) & 5.0 (5.0) & 0.9 (1.0) & \NA \\
 & $\PW + \text{jets}$ &  2.9 (1.6) & 14 (6.8) &  12 (11) & 5.0 (5.0) & 4.8 (5.7) & \NA \\
 & $\PZ/\PGg + \text{jets}$ &  0.7 (3.4) & 9.0 (11) &  12 (11) & 4.5 (4.5) & 5.9 (5.9) & \NA \\
 & $\PV\PV$ &  0.6 (4.4) & 15 (49) &  10 (9.4) & 4.0 (4.0) & 20 (36) & \NA \\
 & QCD multijet &  \NA & \NA &  \NA & 10 (10) & 19 (9.4) & \NA \\
[\cmsTabSkip]
 Tight  & $\mhp=100\GeV$ &  1.2 (1.3) & 2.2 (3.0) &  9.2 (9.2) & 6.1 (6.1) & 1.6 (1.9) & 1.4 (1.8) \\
 & \ttbar &  0.9 (1.0) & 2.7 (3.1) &  9.4 (9.4) & 6.1 (6.1) & 0.4 (0.5) & 1.4 (1.8) \\
 & Single \PQt quark &  0.4 (0.5) & 4.3 (4.5) &  9.8 (9.8) & 5.0 (5.0) & 1.8 (2.1) & \NA \\
 & $\PW + \text{jets}$ &  1.1 (2.8) & 23 (3.4) &  13 (13) & 5.0 (5.0) & 12 (14) & \NA \\
 & $\PZ/\PGg + \text{jets}$ &  3.7 (2.7) & 7.5 (10) &  13 (12) & 4.5 (4.5) & 9.1 (15) & \NA \\
 & $\PV\PV$ &  2.3 (8.9) & 36 (0.3) &  11 (10) & 4.0 (4.0) & 38 (100) & \NA \\
 & QCD multijet &  \NA & \NA &  \NA & 10 (10) & 47 (17) & \NA \\ 
\end{scotch}
}
\end{table*}

The most important sources of uncertainties in terms of impact on the
expected limit on \Bthb for 
$\mhp=100\GeV$, after the individual charm tagging categories and 
the muon and electron channels have been combined, as discussed in 
Section~\ref{s:secResults}, are the lepton selection (3.8\%), QCD 
multijet background estimate (2.4\%), \ttbar cross section (1.9\%), 
and \PQb/\PQc tagging (1.9\%). The effect of each of the remaining 
systematic uncertainties on the expected limit is estimated to be 
less than 0.3\%.

The number of events in the background processes and the corresponding
uncertainty bands shown in Fig.~\ref{fig:mjj_cTagEx} are obtained using a
background-only fit to data. After the fit, several uncertainties (both statistical
and systematic) are significantly anticorrelated, resulting in a reduction in
the overall uncertainty. This is a feature of doing an extended maximum
likelihood fit. The anticorrelations reflect the fact that while our analysis
can constrain the background normalization with the statistical power of the
data, it cannot distinguish as well between different sources which do not
represent independent degrees of freedom in the model. Prior to the fit, as
shown in Table~\ref{tab:sysUnc}, they are either uncorrelated or positively
correlated.

\section{Results}
\label{s:secResults}
After applying all selection requirements, the expected number of 
background events agrees with the data within the uncertainties. 
The absence of a charged Higgs boson signal in the 
data is characterized by setting exclusion limits on the branching 
fraction \Bthb. 
An asymptotic 95\% \CL limit on \Bthb is 
calculated using the \CLs method~\cite{Junk:1999kv,Read:2002hq} with
likelihood ratios~\cite{Cowan:2010js}:
\begin{linenomath}
\begin{equation}
    \widetilde{q}_{x} = -2 \ln \frac{\mathcal{L}(\text{data}|x,
    \hat{\Uptheta}_{x})}{\mathcal{L}(\text{data}|\hat{x},\hat{\Uptheta})}.
\end{equation}  
\end{linenomath}
where the likelihood is defined as 
\begin{linenomath}
\begin{equation}
    \mathcal{L}(\text{data}|x,\Uptheta) = \prod_{j=1}^{3}\prod_{i=1}^{N}\frac{N_{ij}(x, 
    \Uptheta)^{n_{ij}}}{n_{ij}!}\re^{-N_{ij}(x, \Uptheta)} \prod_k
    p(\widetilde{\Uptheta}_k|\Uptheta_k).
\label{eq:likelihood}
\end{equation}
\end{linenomath}

In this equation, $x = \Bthb$ is the parameter of interest, the first product over
$j$ designates the three charm tagging categories, and $i$ runs over the bins of the \mjj distributions shown
in Fig.~\ref{fig:mjj_cTagEx}. For a given mass bin $i$ and charm tagging category $j$, $n_{ij}$ is the
observed number of events in that bin and charm tagging category, and $N_{ij}(\Uptheta)$ is the expected
number of events. The last term is the product over the individual nuisance parameters $k$ of the probability
density function $p(\widetilde{\Uptheta}_k|\Uptheta_k)$, where $\Uptheta_k$ is the value of the nuisance
parameter.  The estimators $\hat{x}$ and $\hat{\Uptheta}$ correspond to the global maximum of the likelihood
defined in Eq.~\ref{eq:likelihood}. The expected number of events $N_{ij}(\Uptheta)$ is given by, in the
presence of signal:
\begin{linenomath}
\ifthenelse{\boolean{cms@external}}
{\begin{multline}
\label{nbsm}
N_{ij}(x,\Uptheta) = 2x(1-x)N_{ij}^{\ttbar \to \PSHp \PWm}(\Uptheta) \\
+(1-x)^2 N_{ij}^{\ttbar\to \PW^{\pm}\PW^{\mp}}(\Uptheta) + 
N_{ij}^{\text{other}}(\Uptheta),
\end{multline}}
{\begin{equation}
\label{nbsm}
N_{ij}(x,\Uptheta) = 2x(1-x)N_{ij}^{\ttbar \to \PSHp \PWm}(\Uptheta) + 
(1-x)^2 N_{ij}^{\ttbar\to \PW^{\pm}\PW^{\mp}}(\Uptheta) + 
N_{ij}^{\text{other}}(\Uptheta),
\end{equation}}
\end{linenomath}
and in the absence:
\begin{linenomath}
   \begin{equation}
       N_{ij}(\Uptheta) = N_{ij}^{\ttbar \to \PW^{\pm}\PW^{\mp}}(\Uptheta) + N_{ij}^{\text{other}}(\Uptheta),
   \end{equation}
   \end{linenomath}
where $N_{ij}^{\ttbar\to \PSHp\PWm}(\Uptheta)$ and $N_{ij}^{\ttbar \to \PW^{\pm}\PW^{\mp}}(\Uptheta)$ are the
number of events from the simulated signal process and the SM \ttbar process, respectively. Both are
normalized to the expected \ttbar cross sections, as described in Section~\ref{s:secDataMC}. The factor of 2
in Eq.~\ref{nbsm} is derived from the assumption that the event yield and \Bthbbar for \PSHm are the same as
those of \PSHp.

\begin{figure*}
    \centering
    \includegraphics[width=0.45\textwidth]{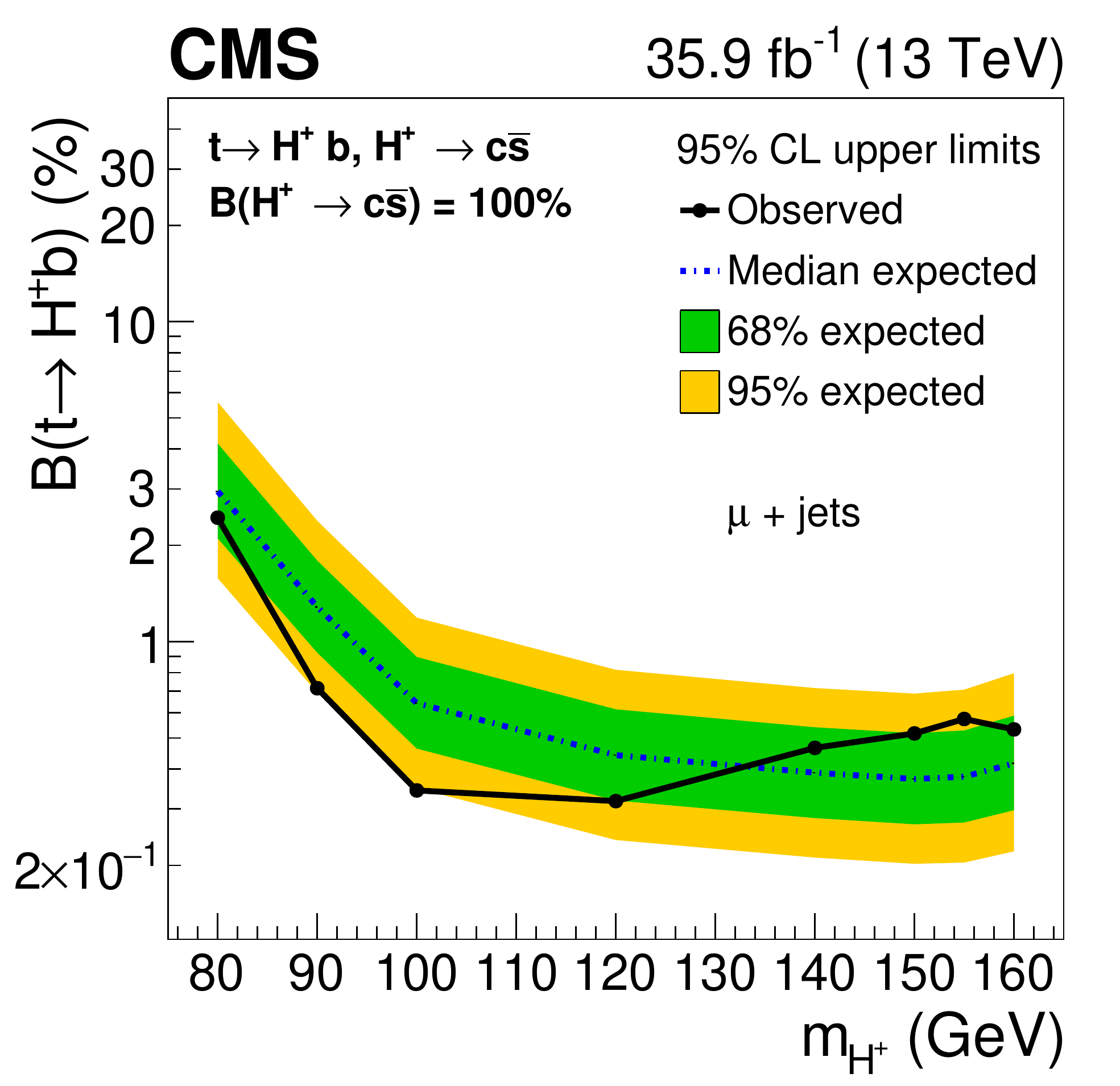}
    \includegraphics[width=0.45\textwidth]{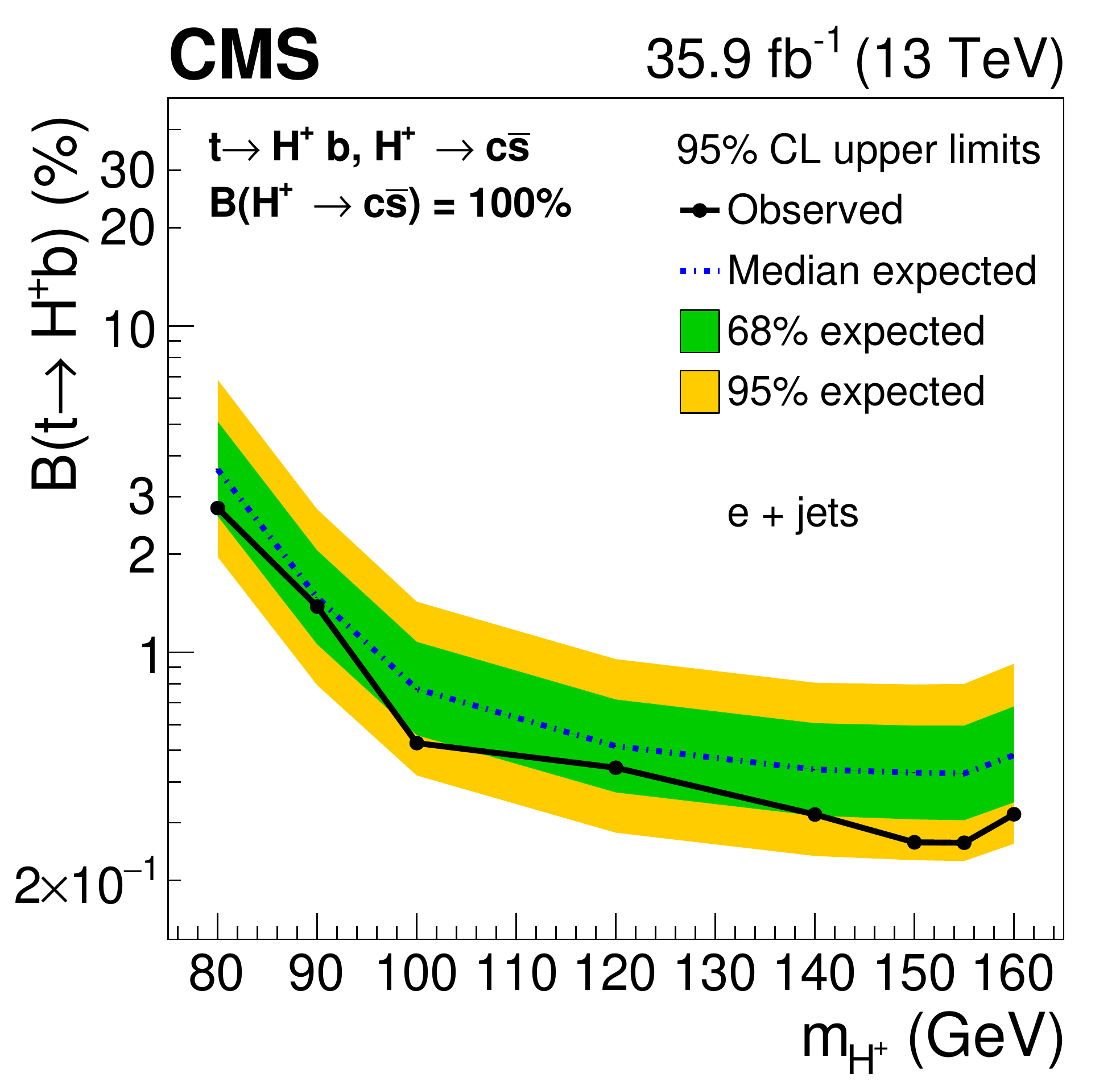}
    \includegraphics[width=0.55\textwidth]{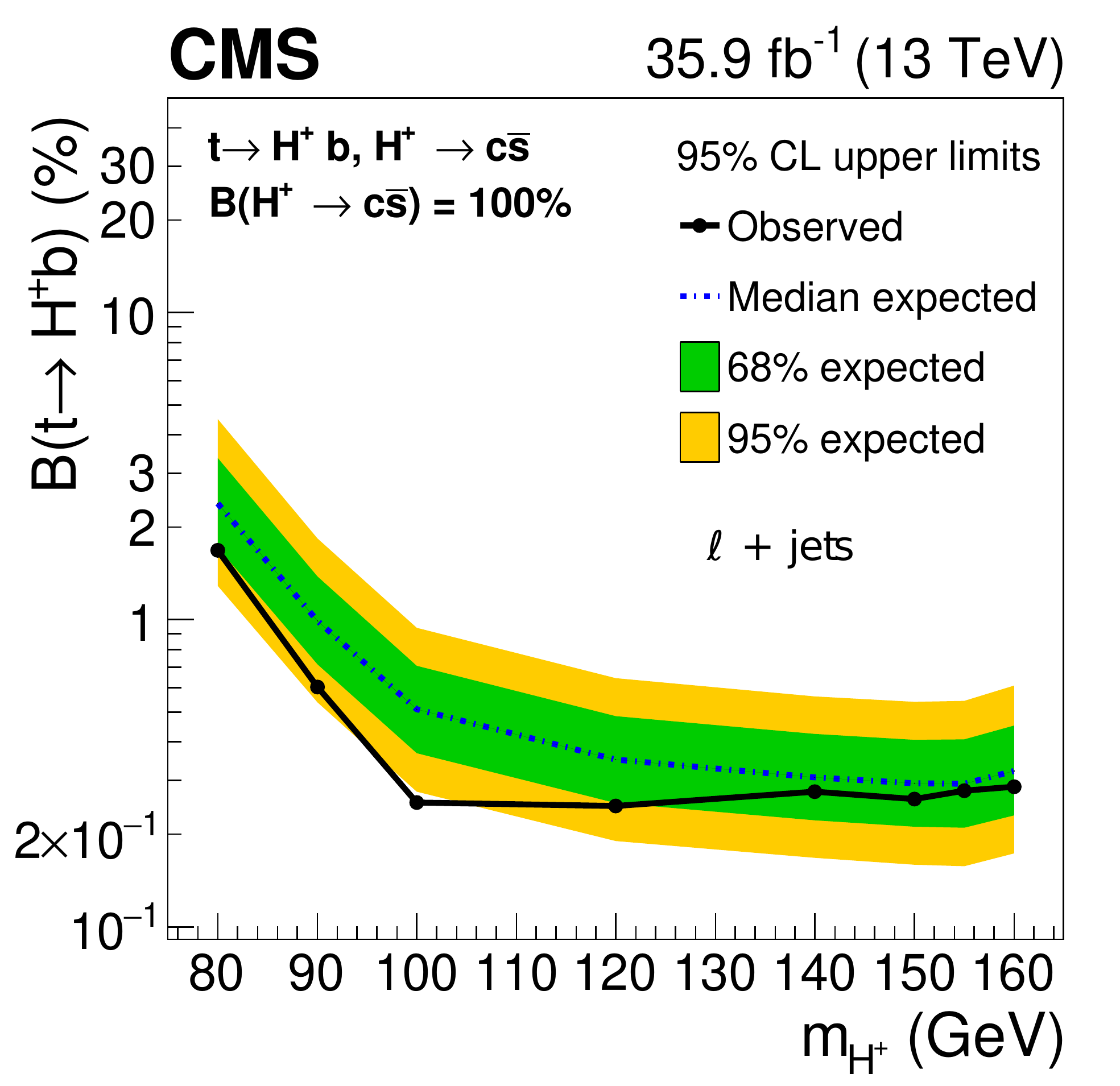}
    \caption{The expected and observed upper limit in \% on
        \Bthb as a function of $\mhp$
        using \mjj after the individual charm tagging categories have
        been combined, for the muon\,+\,jets (upper left) and
        electron\,+\,jets (upper right) channels, and their combination (lower).}
 \label{fig:limitPlot}
\end{figure*}

The exclusion limits on \Bthb as a function of
charged Higgs boson mass using the \mjj distribution in the range 
15--165\GeV and combining
different exclusive event categories based on charm tagging are shown
in Fig.~\ref{fig:limitPlot} and in Tables~\ref{tab:limit_muon_ele} and
\ref{tab:limit_lepton}. Among the individual categories, the expected
limits from the exclusive medium category are most stringent, followed
by those from the exclusive loose and tight categories. By construction,
the exclusion limits on \Bthbbar are the same as those on \Bthb.

\renewcommand{\arraystretch}{1.2}
\begin{table*}[ht]
\centering
\topcaption{Expected and observed 95\% \CL exclusion limits in \% on
    \Bthb in the muon\,+\,jets
    (electron\,+\,jets) channel, after the individual charm tagging categories have
    been combined.}
\label{tab:limit_muon_ele}
\begin{scotch}{ccccccc}
\multirow{2}{*}{$\mhp$ (\GeVns)} & \multicolumn{5}{c}{Expected} & \multirow{2}{*}{Observed} \\
& $-2\sigma$ & $-1\sigma$ & median & $+1\sigma$ & $+2\sigma$ &\\
\hline
80  & 1.58 (1.96) & 2.10 (2.61) & 2.95 (3.63) & 4.16 (5.10) & 5.61 (6.84) & 2.44 (2.77)\\
90  & 0.69 (0.79) & 0.92 (1.06) & 1.28 (1.47) & 1.79 (2.05) & 2.39 (2.74) & 0.72 (1.38)\\
100 & 0.35 (0.42) & 0.46 (0.56) & 0.64 (0.77) & 0.90 (1.08) & 1.19 (1.43) & 0.34 (0.53)\\
120 & 0.24 (0.28) & 0.32 (0.37) & 0.44 (0.52) & 0.61 (0.72) & 0.82 (0.95) & 0.32 (0.44)\\
140 & 0.21 (0.24) & 0.28 (0.32) & 0.39 (0.44) & 0.54 (0.61) & 0.72 (0.81) & 0.47 (0.32)\\
150 & 0.20 (0.23) & 0.27 (0.31) & 0.37 (0.43) & 0.52 (0.60) & 0.69 (0.80) & 0.52 (0.26)\\
155 & 0.20 (0.23) & 0.27 (0.31) & 0.38 (0.42) & 0.53 (0.60) & 0.71 (0.80) & 0.57 (0.26)\\
160 & 0.22 (0.26) & 0.30 (0.35) & 0.42 (0.48) & 0.59 (0.68) & 0.80 (0.92) & 0.53 (0.32)\\
\end{scotch}
\end{table*}

\begin{table}[ht]
\centering
\topcaption{Expected and observed 95\% \CL exclusion limits in \% on
    \Bthb, after the individual charm
    tagging categories and the muon and electron channels have been
    combined.}
\label{tab:limit_lepton}
\begin{scotch}{ccccccc}
\multirow{2}{*}{$\mhp$ (\GeVns)} & \multicolumn{5}{c}{Expected} & \multirow{2}{*}{Observed} \\
& $-2\sigma$ & $-1\sigma$ & median & $+1\sigma$ & $+2\sigma$ & \\
\hline
80  & 1.29 & 1.72 & 2.39 & 3.36 & 4.50 & 1.68\\
90  & 0.54 & 0.72 & 0.99 & 1.38 & 1.84 & 0.60\\
100 & 0.28 & 0.37 & 0.51 & 0.71 & 0.94 & 0.25\\
120 & 0.19 & 0.25 & 0.35 & 0.49 & 0.64 & 0.25\\
140 & 0.17 & 0.22 & 0.31 & 0.42 & 0.56 & 0.28\\
150 & 0.16 & 0.21 & 0.29 & 0.41 & 0.54 & 0.26\\
155 & 0.16 & 0.21 & 0.29 & 0.41 & 0.54 & 0.28\\
160 & 0.17 & 0.23 & 0.32 & 0.45 & 0.61 & 0.29\\
\end{scotch}
\end{table}

\section{Summary}
\label{s:secConcl}
A search for a light charged Higgs boson produced by top quark
decay has been performed in the muon\,+\,jets and electron\,+\,jets
channels at $\sqrt{s} = 13\TeV$, using a data sample corresponding to
an integrated luminosity of 35.9\fbinv. The observed and predicted
number of events from standard model processes are in agreement within 
the uncertainties. An exclusion limit at 95\% confidence level on the 
branching fraction \Bthb has
been computed by assuming $\BHcs = 100\%$.
The observed exclusion limits are in the range, for a charged Higgs
boson mass between 80 and 160\GeV, 2.44--0.32, 2.77--0.26, and
1.68--0.25\% for the muon\,+\,jets, electron\,+\,jets, and the 
combination of the two channels, respectively. These are the first 
results from the LHC at $\sqrt{s}=13\TeV$ for the above final states,
and represent an improvement by a factor of approximately 4 over
the previous results at $\sqrt{s}=8\TeV$.

\begin{acknowledgments}
    We congratulate our colleagues in the CERN accelerator departments for the excellent performance of the LHC and thank the technical and administrative staffs at CERN and at other CMS institutes for their contributions to the success of the CMS effort. In addition, we gratefully acknowledge the computing centers and personnel of the Worldwide LHC Computing Grid for delivering so effectively the computing infrastructure essential to our analyses. Finally, we acknowledge the enduring support for the construction and operation of the LHC and the CMS detector provided by the following funding agencies: BMBWF and FWF (Austria); FNRS and FWO (Belgium); CNPq, CAPES, FAPERJ, FAPERGS, and FAPESP (Brazil); MES (Bulgaria); CERN; CAS, MoST, and NSFC (China); COLCIENCIAS (Colombia); MSES and CSF (Croatia); RPF (Cyprus); SENESCYT (Ecuador); MoER, ERC IUT, PUT and ERDF (Estonia); Academy of Finland, MEC, and HIP (Finland); CEA and CNRS/IN2P3 (France); BMBF, DFG, and HGF (Germany); GSRT (Greece); NKFIA (Hungary); DAE and DST (India); IPM (Iran); SFI (Ireland); INFN (Italy); MSIP and NRF (Republic of Korea); MES (Latvia); LAS (Lithuania); MOE and UM (Malaysia); BUAP, CINVESTAV, CONACYT, LNS, SEP, and UASLP-FAI (Mexico); MOS (Montenegro); MBIE (New Zealand); PAEC (Pakistan); MSHE and NSC (Poland); FCT (Portugal); JINR (Dubna); MON, RosAtom, RAS, RFBR, and NRC KI (Russia); MESTD (Serbia); SEIDI, CPAN, PCTI, and FEDER (Spain); MOSTR (Sri Lanka); Swiss Funding Agencies (Switzerland); MST (Taipei); ThEPCenter, IPST, STAR, and NSTDA (Thailand); TUBITAK and TAEK (Turkey); NASU (Ukraine); STFC (United Kingdom); DOE and NSF (USA). 

    \hyphenation{Rachada-pisek} Individuals have received support from the Marie-Curie program and the European Research Council and Horizon 2020 Grant, contract Nos.\ 675440, 752730, and 765710 (European Union); the Leventis Foundation; the A.P.\ Sloan Foundation; the Alexander von Humboldt Foundation; the Belgian Federal Science Policy Office; the Fonds pour la Formation \`a la Recherche dans l'Industrie et dans l'Agriculture (FRIA-Belgium); the Agentschap voor Innovatie door Wetenschap en Technologie (IWT-Belgium); the F.R.S.-FNRS and FWO (Belgium) under the ``Excellence of Science -- EOS" -- be.h project n.\ 30820817; the Beijing Municipal Science \& Technology Commission, No. Z191100007219010; the Ministry of Education, Youth and Sports (MEYS) of the Czech Republic; the Deutsche Forschungsgemeinschaft (DFG) under Germany's Excellence Strategy -- EXC 2121 ``Quantum Universe" -- 390833306; the Lend\"ulet (``Momentum") Program and the J\'anos Bolyai Research Scholarship of the Hungarian Academy of Sciences, the New National Excellence Program \'UNKP, the NKFIA research grants 123842, 123959, 124845, 124850, 125105, 128713, 128786, and 129058 (Hungary); the Council of Science and Industrial Research, India; the HOMING PLUS program of the Foundation for Polish Science, cofinanced from European Union, Regional Development Fund, the Mobility Plus program of the Ministry of Science and Higher Education, the National Science Center (Poland), contracts Harmonia 2014/14/M/ST2/00428, Opus 2014/13/B/ST2/02543, 2014/15/B/ST2/03998, and 2015/19/B/ST2/02861, Sonata-bis 2012/07/E/ST2/01406; the National Priorities Research Program by Qatar National Research Fund; the Ministry of Science and Education, grant no. 14.W03.31.0026 (Russia); the Tomsk Polytechnic University Competitiveness Enhancement Program and ``Nauka" Project FSWW-2020-0008 (Russia); the Programa Estatal de Fomento de la Investigaci{\'o}n Cient{\'i}fica y T{\'e}cnica de Excelencia Mar\'{\i}a de Maeztu, grant MDM-2015-0509 and the Programa Severo Ochoa del Principado de Asturias; the Thalis and Aristeia programs cofinanced by EU-ESF and the Greek NSRF; the Rachadapisek Sompot Fund for Postdoctoral Fellowship, Chulalongkorn University and the Chulalongkorn Academic into Its 2nd Century Project Advancement Project (Thailand); the Kavli Foundation; the Nvidia Corporation; the SuperMicro Corporation; the Welch Foundation, contract C-1845; and the Weston Havens Foundation (USA). 
\end{acknowledgments}

\bibliography{auto_generated}
\cleardoublepage \appendix\section{The CMS Collaboration \label{app:collab}}\begin{sloppypar}\hyphenpenalty=5000\widowpenalty=500\clubpenalty=5000\vskip\cmsinstskip
\textbf{Yerevan Physics Institute, Yerevan, Armenia}\\*[0pt]
A.M.~Sirunyan$^{\textrm{\dag}}$, A.~Tumasyan
\vskip\cmsinstskip
\textbf{Institut f\"{u}r Hochenergiephysik, Wien, Austria}\\*[0pt]
W.~Adam, F.~Ambrogi, T.~Bergauer, M.~Dragicevic, J.~Er\"{o}, A.~Escalante~Del~Valle, R.~Fr\"{u}hwirth\cmsAuthorMark{1}, M.~Jeitler\cmsAuthorMark{1}, N.~Krammer, L.~Lechner, D.~Liko, T.~Madlener, I.~Mikulec, F.M.~Pitters, N.~Rad, J.~Schieck\cmsAuthorMark{1}, R.~Sch\"{o}fbeck, M.~Spanring, S.~Templ, W.~Waltenberger, C.-E.~Wulz\cmsAuthorMark{1}, M.~Zarucki
% ...
\vskip\cmsinstskip
\dag: Deceased\\
1:  Also at Vienna University of Technology, Vienna, Austria\\
% ...
\end{sloppypar}
\end{document}